\documentclass[twocolumn,twocolappendix]{aastex63}
\newcommand{\bvec}[1]{\mbox{\boldmath $#1$}}
\begin{document}

\title{Particle Acceleration by Pickup Process Upstream of Relativistic Shocks}

\author{Masanori Iwamoto}
\affiliation{Faculty of Engineering Sciences, Kyushu University,
6-1, Kasuga-koen, Kasuga, Fukuoka, 816-8580, Japan}

\author{Takanobu Amano}
\affiliation{Department of Earth and Planetary Science, University of Tokyo,
 7-3-1 Hongo, Bunkyo-ku, Tokyo 113-0033, Japan}

\author{Yosuke Matsumoto}
\affiliation{Department of Physics, Chiba University, 1-33 Yayoi, Inage-ku,
 Chiba, Chiba 263-8522, Japan}

\author{Shuichi Matsukiyo}
\affiliation{Faculty of Engineering Sciences, Kyushu University,
6-1, Kasuga-koen, Kasuga, Fukuoka, 816-8580, Japan}

\author{Masahiro Hoshino}
\affiliation{Department of Earth and Planetary Science, University of Tokyo,
7-3-1 Hongo, Bunkyo-ku, Tokyo 113-0033, Japan}

\correspondingauthor{Masanori Iwamoto}
\email{iwamoto@esst.kyushu-u.ac.jp}

\accepted{ApJ}

\begin{abstract}

  Particle acceleration at magnetized purely perpendicular
  relativistic shocks in electron-ion plasmas are studied by means of
  two-dimensional particle-in-cell simulations. Magnetized shocks with the
  upstream bulk Lorentz factor $\gamma_1 \gg 1$ are known to
  emit intense electromagnetic waves from the shock front,
  which induce electrostatic plasma waves (wakefield) and transverse
  filamentary structures in the upstream region via the stimulated/induced
  Raman scattering and the filamentation instability, respectively. The
  wakefield and filaments inject a fraction of incoming
  particles into a particle acceleration process, in
  which particles are once decoupled from the upstream bulk flow by
  the wakefield, and are piked up again by the flow. The picked-up
  particles are accelerated by the motional electric field.
  The maximum attainable Lorentz factor is estimated as
  $\gamma_{max,e} \sim \alpha\gamma_1^3$ for electrons and
  $\gamma_{max,i} \sim (1+m_e\gamma_1/m_i)\gamma_1^2$ for ions, where
  $\alpha \sim 10$ is determined from our simulation results.
  $\alpha$ can increase up to $\gamma_1$ for weakly magnetized
  shock if $\gamma_1$ is sufficiently large.
  This result indicates that highly relativistic astrophysical shocks such
  as external shocks of gamma-ray bursts can be an efficient particle
  accelerator.

\end{abstract}

\keywords{Shocks --- Plasma physics--- Cosmic rays --- High energy astrophysics}

\section{Introduction}\label{sec:intro}

  Particle acceleration is a ubiquitous physical process in the universe.
  The nonthermal emission spectra of high-energy astrophysical objects such
  as jets from  active galactic nuclei (AGNs) and gamma-ray bursts (GRBs)
  are generally modeled as the synchrotron and inverse Compton emission of
  relativistic electrons \cite[see, e.g.,][]{Piran2005,Blandford2019}.
  AGN jets and GRBs are often invoked for the source of ultra-high-energy
  cosmic rays (UHECRs) with energies beyond $10^{18}$ eV
  \cite[e.g.,][]{Hillas1984}. Although
  the origin of UHECRs is still unknown, recent observations favor
  the extragalactic origin \citep{Aab2018,Aartsen2018}.
  Such astrophysical objects are usually associated with relativistic shocks
  as a consequence of interaction between jets and interstellar medium.
  Relativistic shocks are assumed to be an efficient particle accelerator.

  Coherent emission of electromagnetic waves
  from the shock front is intrinsic to relativistic
  shocks, which has been confirmed by one-dimensional
  \cite[1D;][]{Langdon1988,Gallant1992,Hoshino1992,Amato2006,Plotnikov2019},
  two-dimensional
  \cite[2D;][]{Iwamoto2017,Iwamoto2018,Plotnikov2018,Babul2020},
  and tree-dimensional \cite[3D;][]{Sironi2021}
  particle-in-cell (PIC) simulations in pair plasmas.
  It results from the synchrotron maser instability (SMI)
  in the shock-transition, which is driven by electrons reflected off the
  shock-compressed magnetic field \citep{Hoshino1991}. The excited
  electromagnetic waves whose group velocities are faster than the shock
  can propagate thorough the upstream plasmas as precursor
  waves. The SMI, which is called cyclotron maser instability in weakly
  relativistic context, is also known as the emission mechanism of
  coherent radio sources such as auroral kilometric radiation in the Earth
  and Jovian decametric radiation \citep[see, e.g.,][]{Melrose2017}.
  Recently, some models of fast radio burst based on the coherent emission
  from relativistic shock via the SMI have been proposed
  \cite[e.g.,][]{Lyubarsky2014,Beloborodov2017,Metzger2019,Beloborodov2020,
  Margalit2020} and the SMI in the context of relativistic shocks attracts
  more attention from astrophysics. The precursor waves excited by the
  SMI can be strong enough to induce filamentation instability (FI)
  which is a transverse self-modulation of an intense electromagnetic wave
  \citep{Kaw1973,Drake1974,Max1974,Sobacchi2020}.
  The previous multidimensional simulations indeed demonstrated
  that the intense electromagnetic waves propagating through upstream
  plasma induce the transverse filamentary structures. The
  nonlinear effect of the electromagnetic waves plays a significant role
  in astrophysical plasmas \cite[see also][]{Lyubarsky2018,Lyubarsky2019}.

  In electron-ion plasmas, the nonlinear interactions between the precursor
  waves and the upstream plasmas become more complicated. The
  stimulated/induced Raman scattering (SRS), which is a parametric
  decay of an intense electromagnetic wave into an electrostatic plasma wave
  such as a Langmuir wave, can work in addition to the FI
  \cite[see, e.g.,][]{Kruer1988,Lyubarsky2008}.
  This plasma wave is conventionally called wakefield
  and the concept of the direct particle acceleration by the wakefield via
  the Landau resonance, which is so-called wakefield acceleration (WFA),
  is first proposed in the study of laboratory plasmas \citep{Tajima1979}.
  The application of the WFA to the UHECR acceleration is
  discussed later in the context of astrophysics
  \citep[e.g.,][]{Chen2002,Arons2003,Murase2009,Ebisuzaki2014,Ebisuzaki2021}
  and laboratory plasmas
  \citep[e.g.,][]{Kuramitsu2008,Kuramitsu2011a,Kuramitsu2011b,Kuramitsu2012,
  Liu2017,Liu2018,Liu2019}.
  In relativistic shocks, \cite{Lyubarsky2006} first found the
  wakefield excitation via the SRS using 1D PIC simulations and recent 2D
  PIC simulations \citep{Sironi2011,Ligorini2021a,Ligorini2021b} confirmed
  that. Furthermore, 1D PIC simulations by
  \cite{Hoshino2008} demonstrated that nonthermal
  electrons and ions are generated in the upstream.
  Although our high-resolution 2D PIC simulations \citep{Iwamoto2019}
  showed that this particle acceleration associated with the wakefield
  operates even in a 2D system, the detailed acceleration mechanism was not
  fully understood because it has some different aspects from the
  standard WFA in laboratory plasmas.

  In this work, we investigate the acceleration mechanism in more detail and
  show that the energetic particles are generated by a pickup
  process, where some incoming particles are decoupled once from the
  upstream bulk flow by the wakefield
  and then accelerated by the motional electric field
  after picked up by the flow. In an ideal case, the maximum Lorentz
  factor may reach $\gamma_{max,e} \sim \alpha\gamma_1^3$
  for electrons and $\gamma_{max,i} \sim (1+m_e\gamma_1/m_i)\gamma_1^2$ for
  ions, where $\alpha \sim 10$ is the factor determined from our simulation
  results and $\gamma_1$ is the upstream bulk Lorentz factor.
  Although the observed Lorentz factor is smaller than this
  theoretical estimate due to the limitation of the simulation time,
  the partial trajectories of energetic particles are well-described by
  the pickup process. This efficient acceleration
  process may operate in highly relativistic astrophysical shocks.

\section{Simulation Setup}\label{sec:setup}

\begin{figure}[htb!]
   \plotone{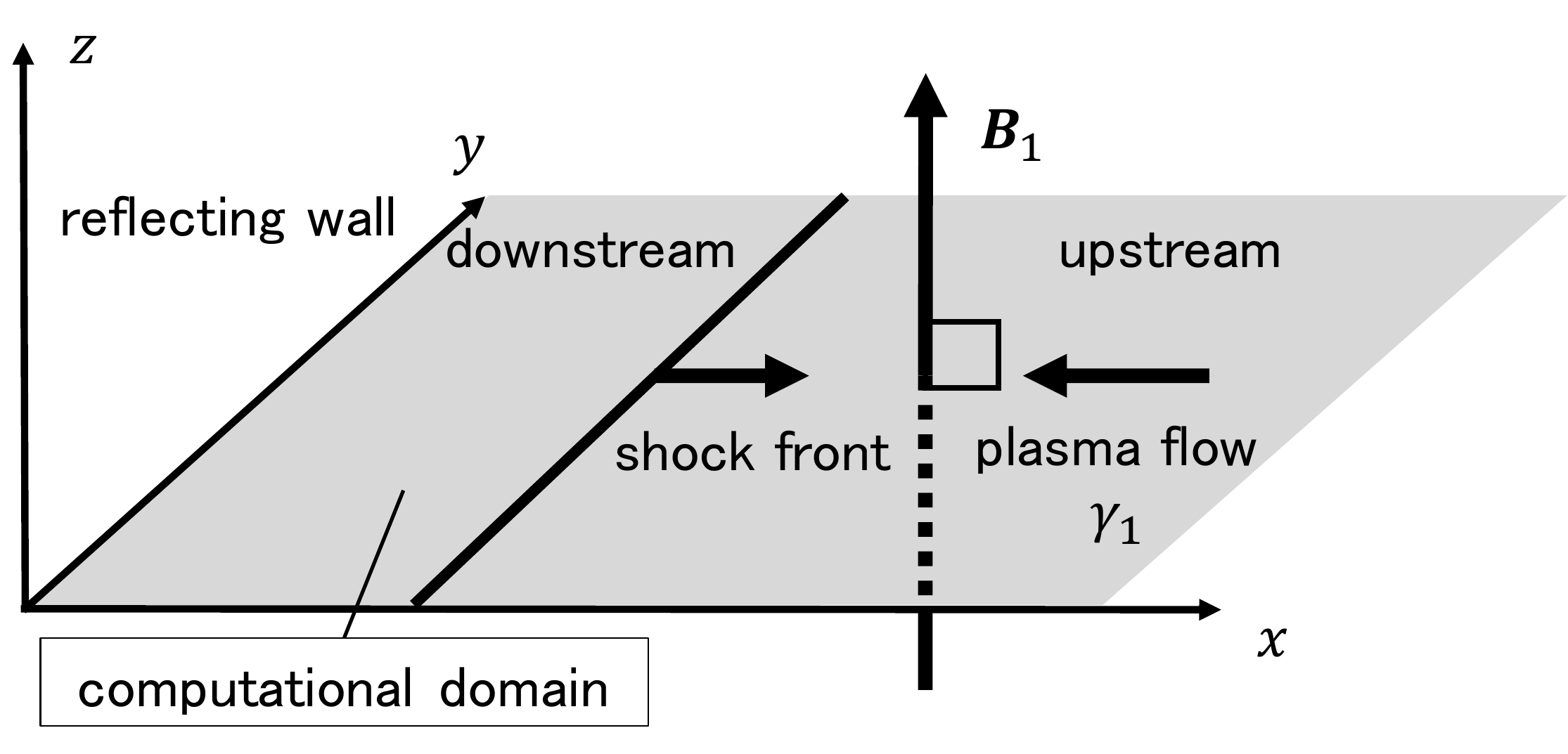}
   \caption{Coordinate system and simulation geometry.}
   \label{fig:geo}
  \end{figure}

  We perform 2D simulations of perpendicular relativistic shocks
  in electron-ion plasmas by using a fully kinetic electromagnetic PIC code
  \citep{Matsumoto2013, Matsumoto2015}, which suppresses the
  numerical Cherenkov instability by choosing a magic CFL number
  and enables accurate and stable calculation \citep{Ikeya2015}.
  Our basic configuration is illustrated in Figure \ref{fig:geo}.
  We consider a rectangular computational domain in the $x$--$y$ plane
  with the periodic boundary condition applied in the $y$ direction.
  The number of grids in each direction is
  $N_x \times N_y = 80,000 \times 1,536$.
  The number of particles per cell per species in the upstream and the gird
  size are respectively set as  $N_1 \Delta x^2 = 64$ and
  $\Delta x/(c/\omega_{pe})= 1/40$, which are motivated by the
  numerical convergence study of our previous simulations
  \citep[see][Appendix A]{Iwamoto2017}. $c$ is the speed of light and
  $\omega_{pe}$ is the proper electron plasma frequency:
  \begin{equation}
    \omega_{pe} = \sqrt{\frac{4 \pi N_1 e^2}{\gamma_1 m_e}}.
  \end{equation}
  The time step is automatically determined as $\omega_{pe} \Delta t = 1/40$
  because the magic CFL number is $c\Delta t/\Delta x = 1$ for our implicit
  Maxwell solver. Note that the implicit Maxwell solver is not restricted
  by the CFL number and thus $c\Delta t/\Delta x = 1$ is numerically
  stable. A cold ion-electron flow with the ion-to-electron mass ratio
  $m_i/m_e=50$ is injected from the right-hand boundary and
  propagating $-x$ direction with the bulk Lorentz factor $\gamma_1$.
  Our shock simulations are performed for the two cases: $\gamma_1=40$ and $100$.
  The incoming particles are reflected off at the left-hand conducting-wall
  boundary and trigger the shock propagating $+x$ direction.
  Our simulation frame corresponds to the downstream rest frame.
  We focus on purely perpendicular shocks and the upstream ambient magnetic
  field $B_1$ is in the $z$ direction.

  The basic structure and coherent emission of relativistic magnetized
  shocks are well-characterized by the ratio of the Poynting flux to the
  upstream bulk kinetic energy flux:
  \begin{equation}
    \sigma_s = \frac{B_1^2}{4\pi\gamma_1 N_1m_sc^2}
    =\frac{\omega_{cs}^2}{\omega_{ps}^2},
  \end{equation}
  where subscript $s=e,i$ represents the particle species and
  $\omega_{cs}$ is the relativistic cyclotron frequency:
  \begin{equation}
     \omega_{cs} = \frac{eB_1}{\gamma_1m_sc}.
  \end{equation}
  We use fixed values of $\sigma_i = 0.1$
  and $\sigma_e = (m_i/m_e)\sigma_i = 5$ throughout this study.

\section{Shock Structure}\label{sec:shock}

\begin{figure*}[htb!]
   \plottwo{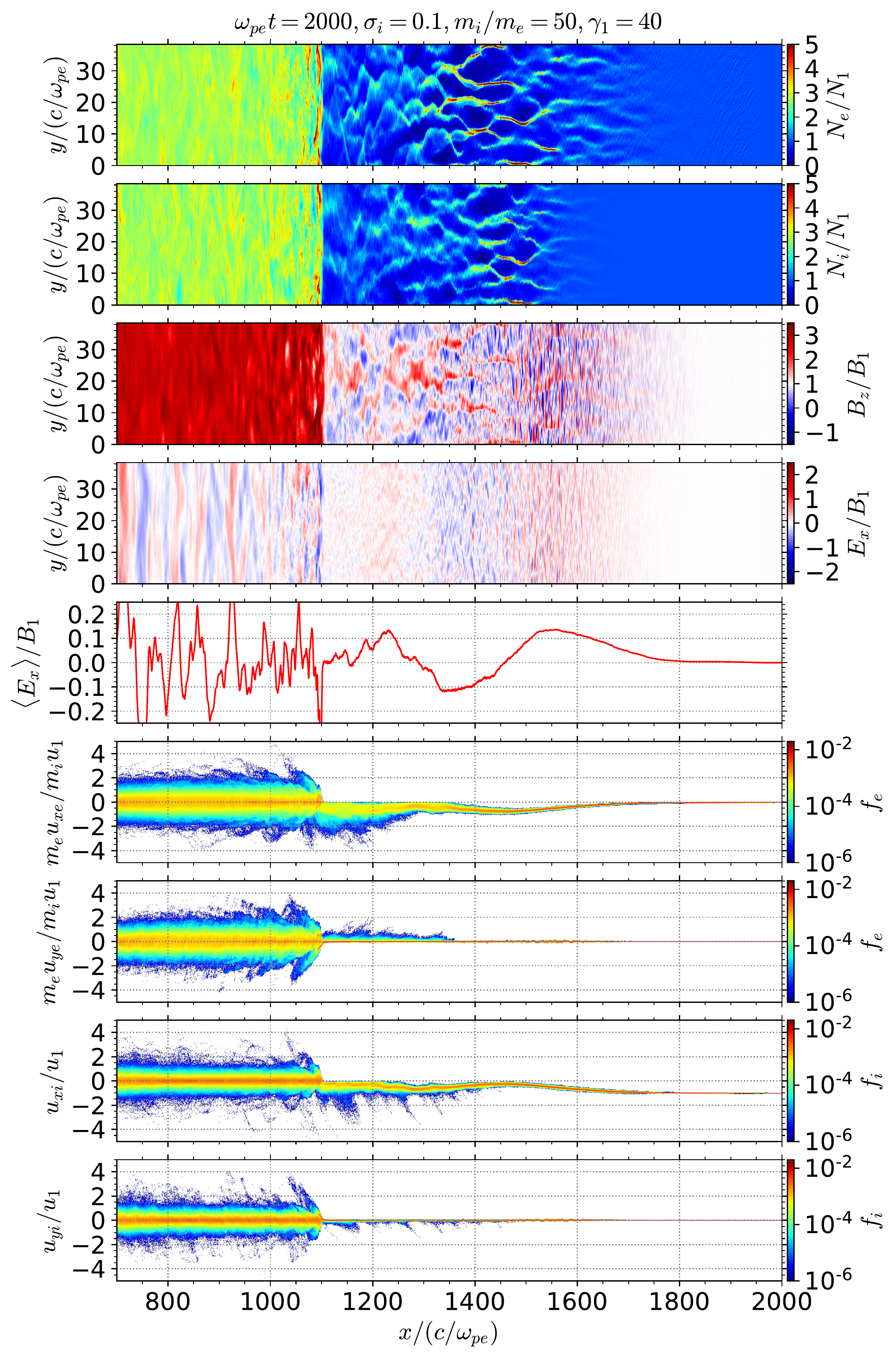}{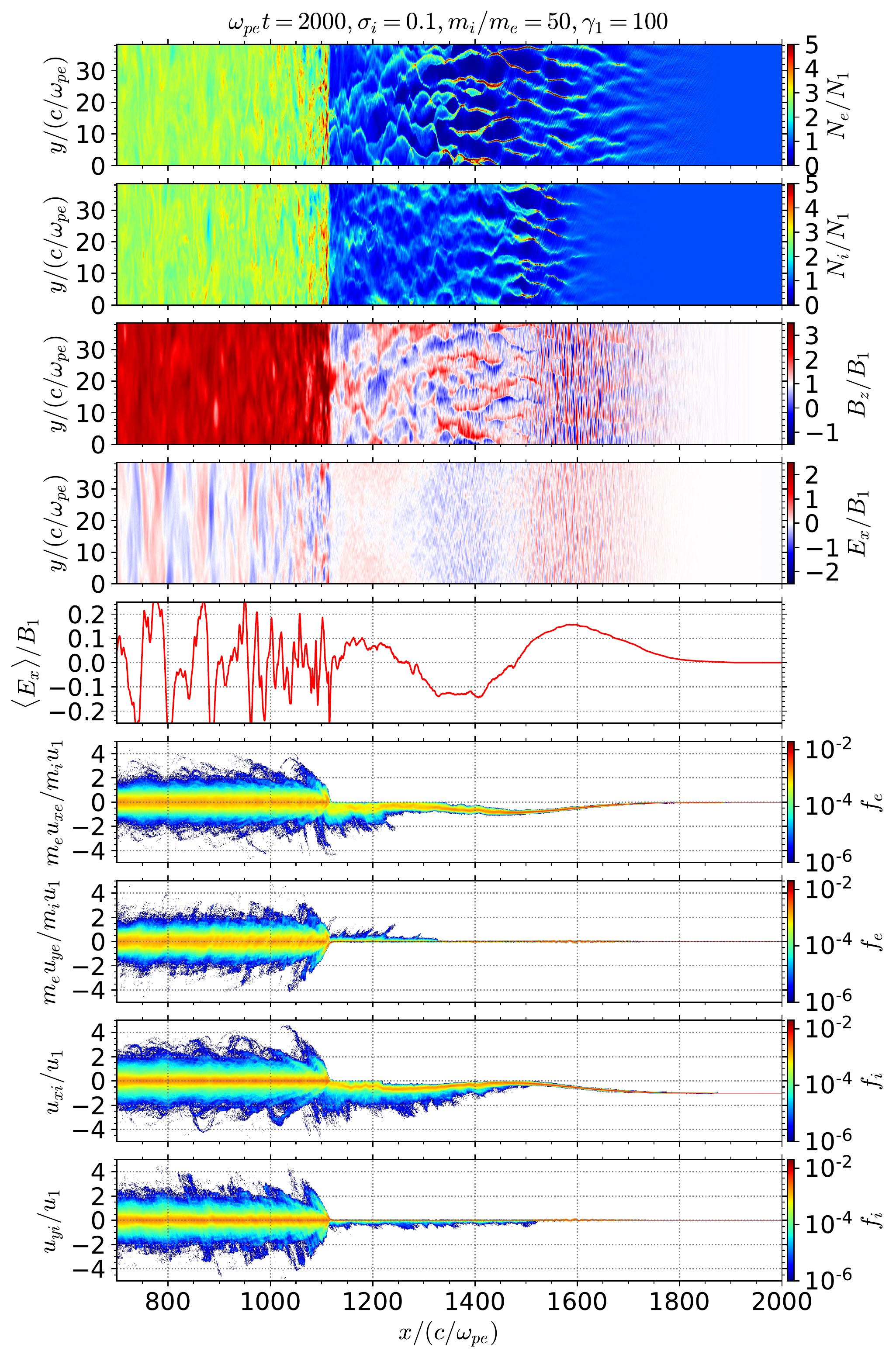}
   \caption{Shock structures at $\omega_{pe}t=2000$ with $\gamma_1=40$
   (left) and $100$ (right). The electron number density $N_e$,
   the ion number density $N_i$, the $z$ component of the magnetic field $B_z$,
   the $x$ component of the electric field $E_x$, the transversely averaged
   electric field $\langle E_x \rangle$, and the phase space densities in the
   $x$--$u_{xs}$ and $x$--$u_{ys}$ are shown.}
   \label{fig:shock}
  \end{figure*}

  Figure \ref{fig:shock} shows the global shock structures at
  $\omega_{pe}t=2000$ in the case of $\gamma_1=40$ (left) and $100$
  (right). The electron number density $N_e$, the ion number density
  $N_i$, the out-of-plane magnetic field $B_z$,
  the longitudinal electric field $E_x$, the longitudinal electric field averaged
  over the $y$ direction $\langle E_x \rangle$, and the phase space densities in
  $x$--$u_{xs}$ and $x$--$u_{ys}$ for both electrons and ions are shown from top
  to bottom.  All quantities are normalized by the corresponding upstream values
  and the electron four velocity $\bvec{u_e}=\gamma_e\bvec{\beta_e}$ are scaled
  by the mass ratio $m_e/m_i$.
  The global structures are similar to each other and show no clear $\gamma_1$
  dependence.

  The clear transverse density filaments are observed in the upstream
  region as in the case of pair plasmas
  \citep{Iwamoto2017,Iwamoto2018,Plotnikov2018,Babul2020,Sironi2021}.
  One might think that the filaments are attributed to
  the Weibel instability \citep{Weibel1959,Fried1959}.
  However, it cannot arise for such high magnetization $\sigma_i=0.1$
  \cite[see, e.g.,][]{Spitkovsky2005,Sironi2011,Sironi2013}.
  We think that the large-amplitude precursor waves, which
  are clearly seen in $B_z$, induce the FI \citep{Kaw1973,Drake1974,Max1974,Sobacchi2020} and create
  the filaments. It requires intense electromagnetic pump waves.
  Actually, the precursor wave amplitude is large in the sense that
  the wave strength parameter $a$ is greater than unity
  \citep{Iwamoto2017,Plotnikov2019}:
  \begin{equation}
    \label{eq:a}
    a = \frac{e\delta E}{m_ec\omega} \sim \gamma_1\sqrt{\epsilon_p} > 1,
  \end{equation}
  for  $\gamma_1 \gg 1$.
  Here $\delta E$ is the wave amplitude, $\omega$ is the wave frequency,
  and $\epsilon_p = \delta B^2/4\pi\gamma_1N_1m_ec^2$
  is the normalized precursor wave energy. $\epsilon_p \sim 1$ for
  $\sigma_i=0.1$ \citep{Iwamoto2019}
  and thus $a \sim \gamma_1 \gg 1$. Therefore,
  the precursor waves are subject to the FI. The FI is a nonlinear wave-wave
  interaction and a kind of the parametric decay
  instability. An intense electromagnetic pump wave with the wave number
  $\bvec{k_0}$ parametrically decays into a compressional wave such as a
  magnetosonic wave with the wave number $\bvec{k_1}$ and two
  forward-scattered electromagnetic waves with the wave number
  $\bvec{k_2} = \bvec{k_0} \pm \bvec{k_1}$.
  The beating of these electromagnetic waves operates when $k_1 \ll k_0$ is
  satisfied. Consequently, the amplitude modulation is induced, which
  in turn enhances the compressional wave via the ponderomotive force.
  This self-modulation can occur in the direction perpendicular to the pump
  wave ($\bvec{k_1} \perp \bvec{k_0}$), resulting in the filaments. Note
  that this simulation frame corresponds to the downstream rest frame. The
  filaments are elongated in the $x$ direction due to
  the Lorentz boost and become more prominent.

  The large-scale electrostatic wave (i.e., wakefield) is generated by the
  SRS in the upstream. The SRS is a kind of the parametric decay
  instability as well. An intense electromagnetic pump with
  $\bvec{\tilde{k}_0}$ parametrically decays into
  the wakefield with $\bvec{\tilde{k}_1}$ and a scattered
  electromagnetic wave with $\bvec{\tilde{k}_2} = \bvec{\tilde{k}_0} - \bvec{\tilde{k}_1}$
  in the linear phase of the SRS \citep{Hoshino2008}.
  The sinusoidal wakfield in the far upstream region
  is induced in this linear phase and particles are merely oscillating as can be
  seen in the phase space plots. The wakefield gradually get turbulent, indicating
  the SRS enter the nonlinear phase. Both electrons and ions are strongly
  accelerated/heated near the shock front.
  Note that the Lorentz transformation from the proper frame into the simulation
  frame increases the thermal spread only in the $u_x$
  direction. The phase space plots
  $x$--$u_{ys}$ show that electrons (ions) are preferentially accelerated
  in the $+y$ ($-y$) direction. This is because particles are mainly accelerated
  by the motional electric field $E_y = -\beta_1B_1$ as discussed in Section
  \ref{sec:acc}.
  These features are consistent with the previous 1D simulation by
  \cite{Hoshino2008}.

\section{Particle energy spectra}\label{sec:spectra}

\begin{figure*}[htb!]
  \gridline{\fig{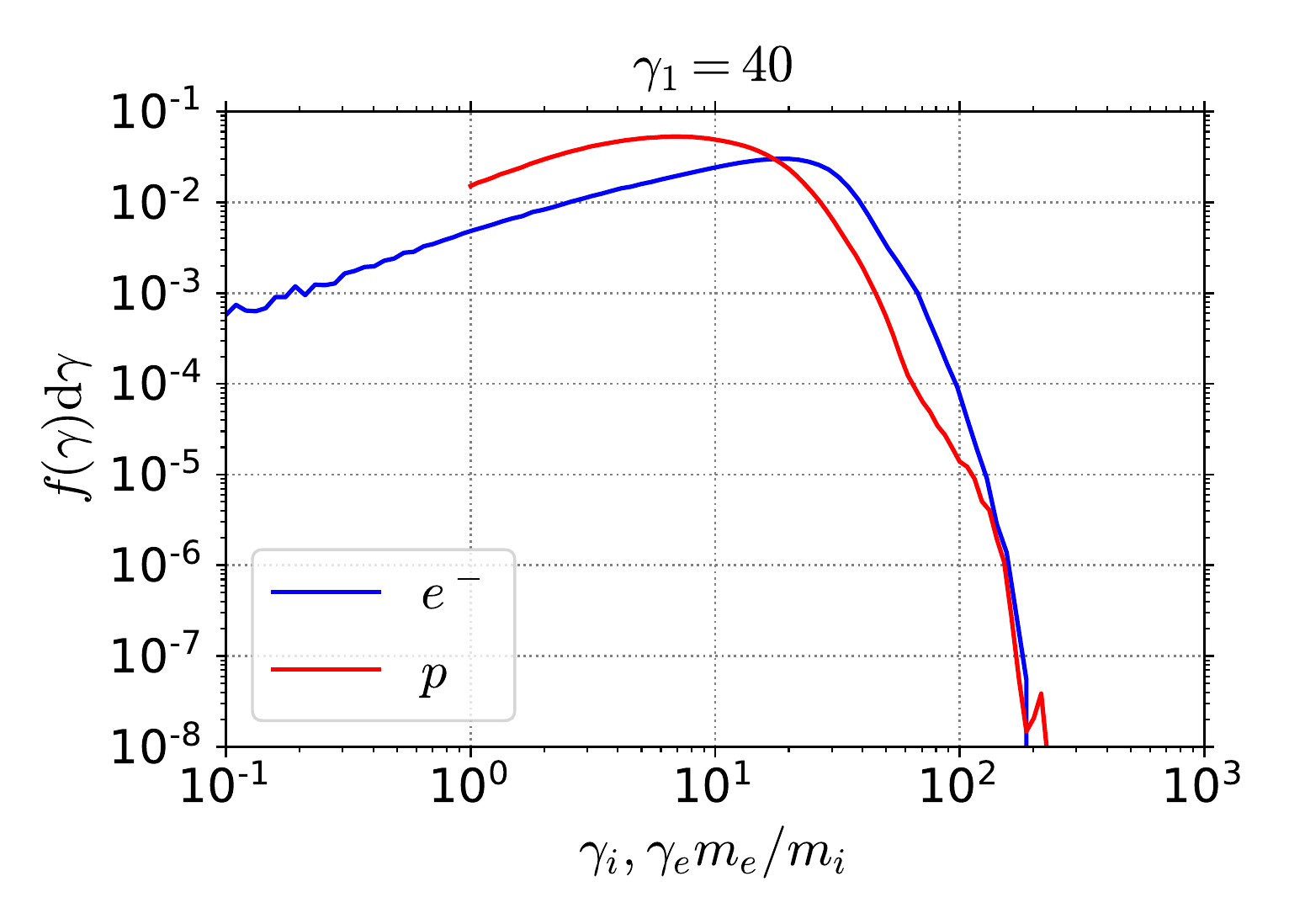}{0.3\textwidth}
            {(a) $700 \le x/(c/\omega_{pe}) \le 1050$}
            \fig{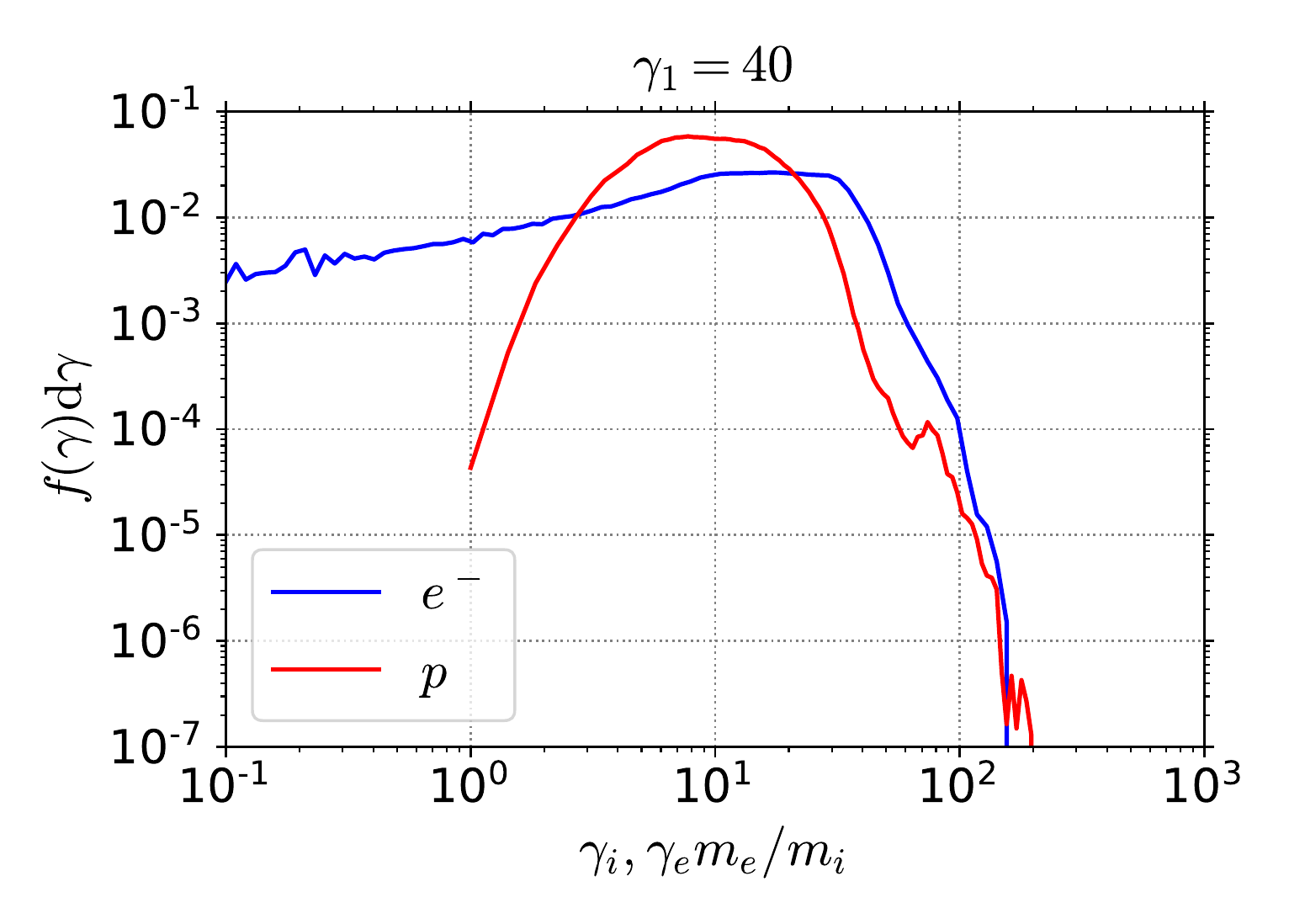}{0.3\textwidth}
            {(b) $1120 \le x/(c/\omega_{pe}) \le 1200$}
            \fig{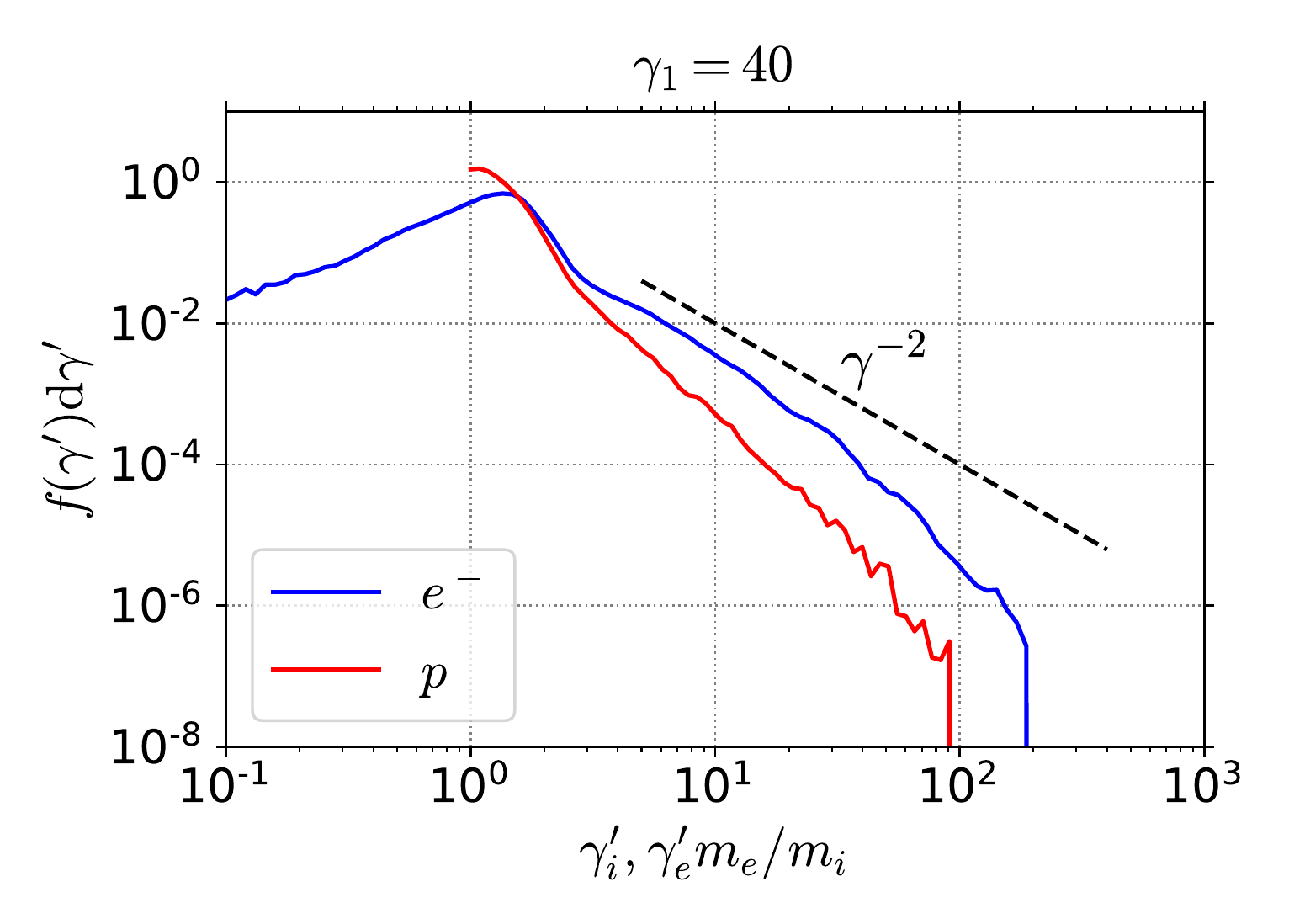}{0.3\textwidth}
            {(c) measured in the proper frame}
            }
  \gridline{\fig{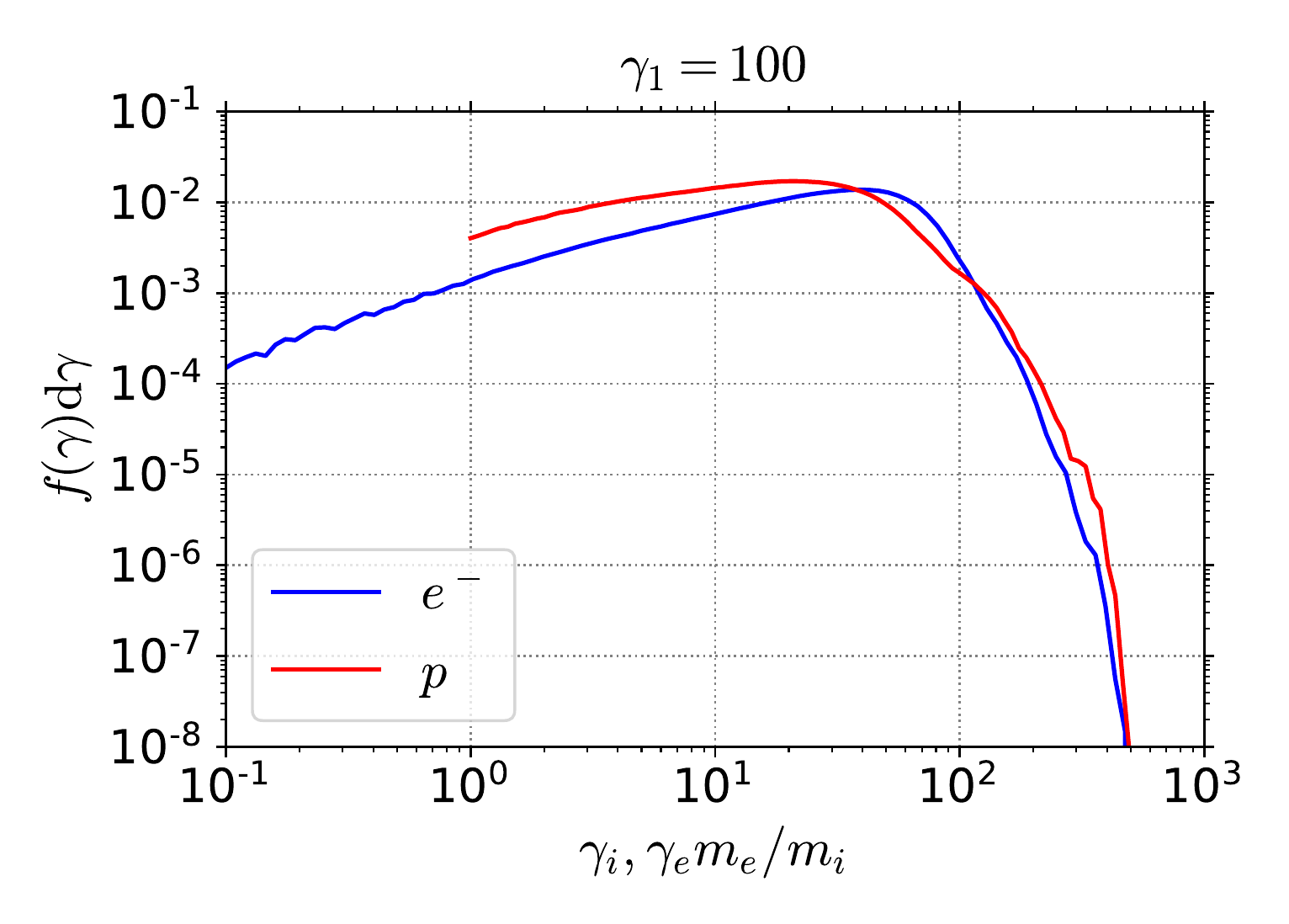}{0.3\textwidth}
            {(d) $700 \le x/(c/\omega_{pe}) \le 1050$}
            \fig{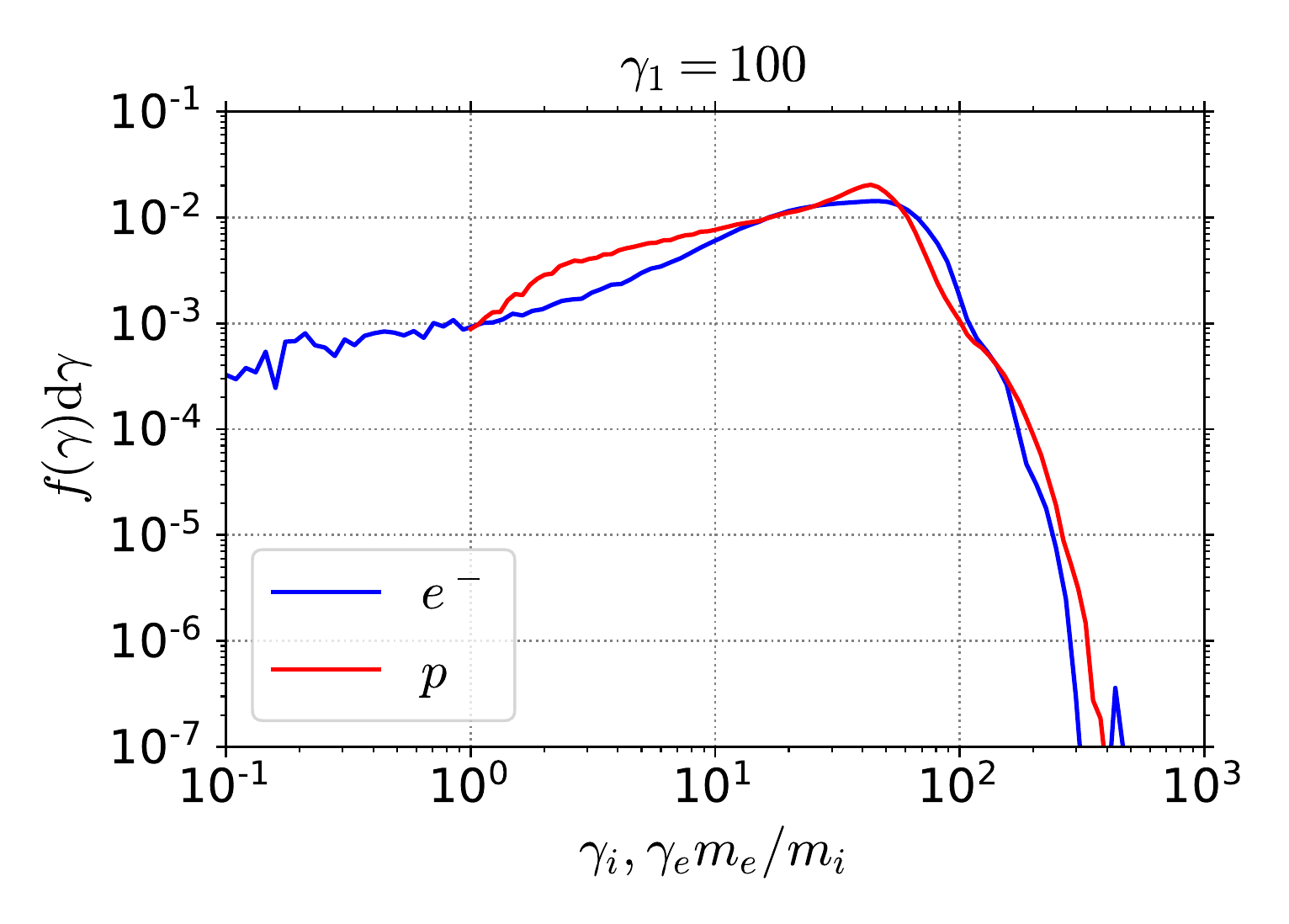}{0.3\textwidth}
            {(e) $1120 \le x/(c/\omega_{pe}) \le 1200$}
            \fig{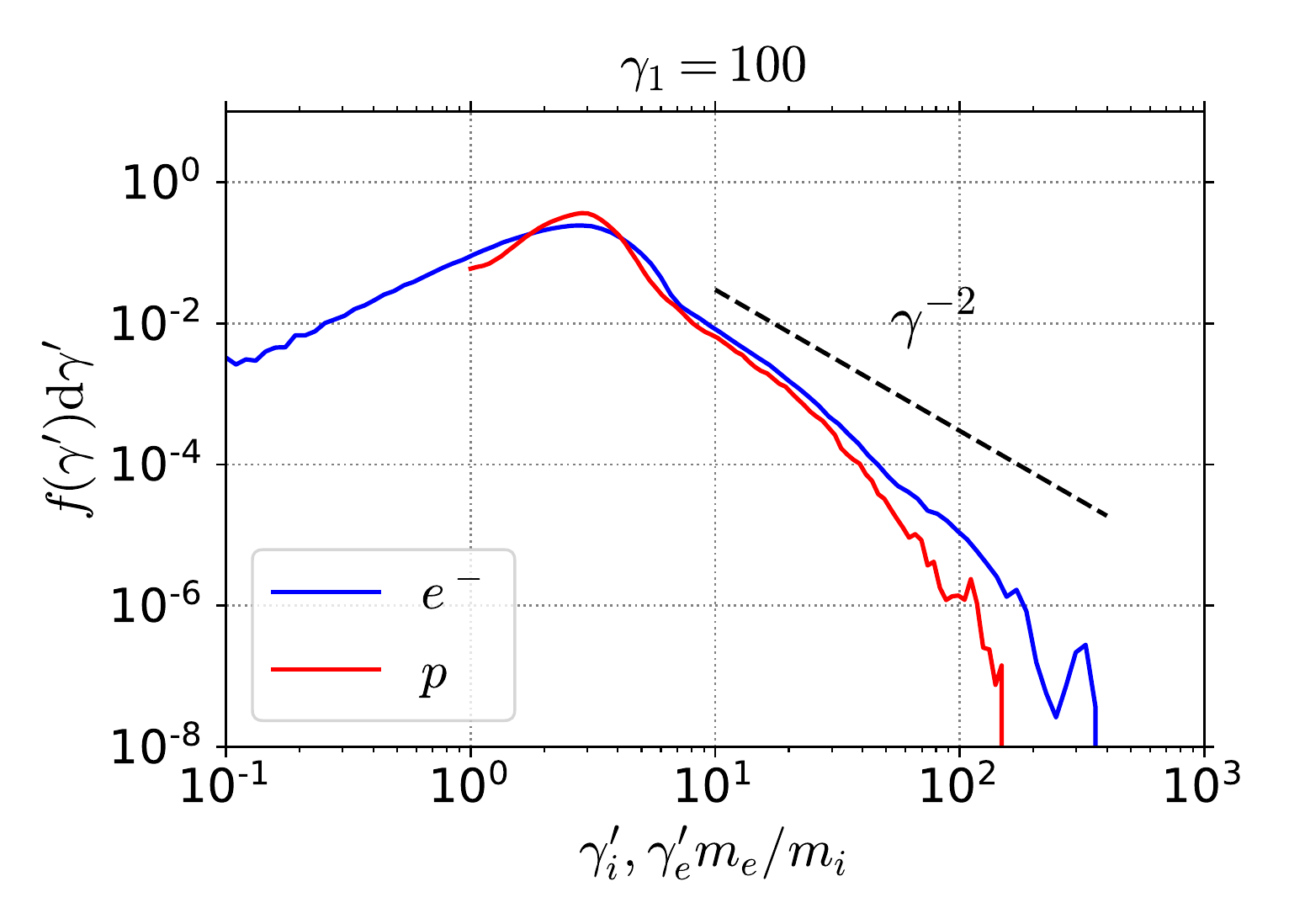}{0.3\textwidth}
            {(f) measured in the proper frame}
            }
  \caption{Energy spectra of electrons (blue) and ions (red) for
  $\gamma_1=40$ (top panels) and $\gamma_1=100$ (bottom panels).
  The downstream spectra (left panels), the upstream spectra
  (middle panels), and the upstream spectra measured in the proper frame
  (right panels) are shown.}
  \label{fig:spectra}
  \end{figure*}

  Figure \ref{fig:spectra} shows energy spectra of electrons (blue)
  and ions (red) at $\omega_{pe}t = 2000$
  for $\gamma_1=40$ (top panels) and $\gamma_1=100$ (bottom panels),
  which are normalized as follows:
  \begin{equation}
    \int_1^{\infty}f_s{\rm d}\gamma =1.
  \end{equation}
  The electron Lorentz factor is scaled by the mass ratio $m_e/m_i$.
  The spectra (a), (b), (d), and (e) are measured in the simulation frame.
  No clear nonthermal tail is seen in the downstream energy spectra (a) and (d)
  in the range $700 \le x/(c/\omega_{pe}) \le 1050$.
  As can be seen in the near-upstream energy spectra (b) and (e)
  in the range $ 1120 \le x/(c/\omega_{pe}) \le 1200$,
  an energy equipartition between electrons and ions is already achieved
  in the upstream due to the electron-ion coupling via the wakefield
  \citep{Lyubarsky2006,Hoshino2008,Iwamoto2019}.
  The typical Lorentz factor can be estimated as
  $\gamma_i = \gamma_em_e/m_i \sim \gamma_1/2$ and
  the peaks of the spectra (a), (b), (d), and (e)
  are roughly consistent with this estimate.

  We also show the upstream spectra measured in the plasma rest frame (c) and (f).
  The prime indicates physical quantities measured in the proper frame.
  We determined the mean bulk velocity in the region:
  $1120 \le x/(c/\omega_{pe}) \le 1200$ and performed Lorentz transformation
  into the plasma rest frame.
  A power-law distribution $\propto \gamma^{-2}$, which is reported
  in the previous laser plasma experiments and simulations of the WFA
  \citep{Kuramitsu2008,Kuramitsu2011a,Kuramitsu2011b,Kuramitsu2012,
    Liu2017,Liu2018,Liu2019}, is also shown in black dashed lines
  for comparison. The nonthermal electrons and ions are clearly visible
  for both $\gamma_1$. These particles are generated by the pickup process as
  discussed in Section \ref{sec:acc}

\section{Acceleration Mechanism}\label{sec:acc}
\subsection{Electron acceleration}\label{subsec:e}

\begin{figure}[htb!]
 \plotone{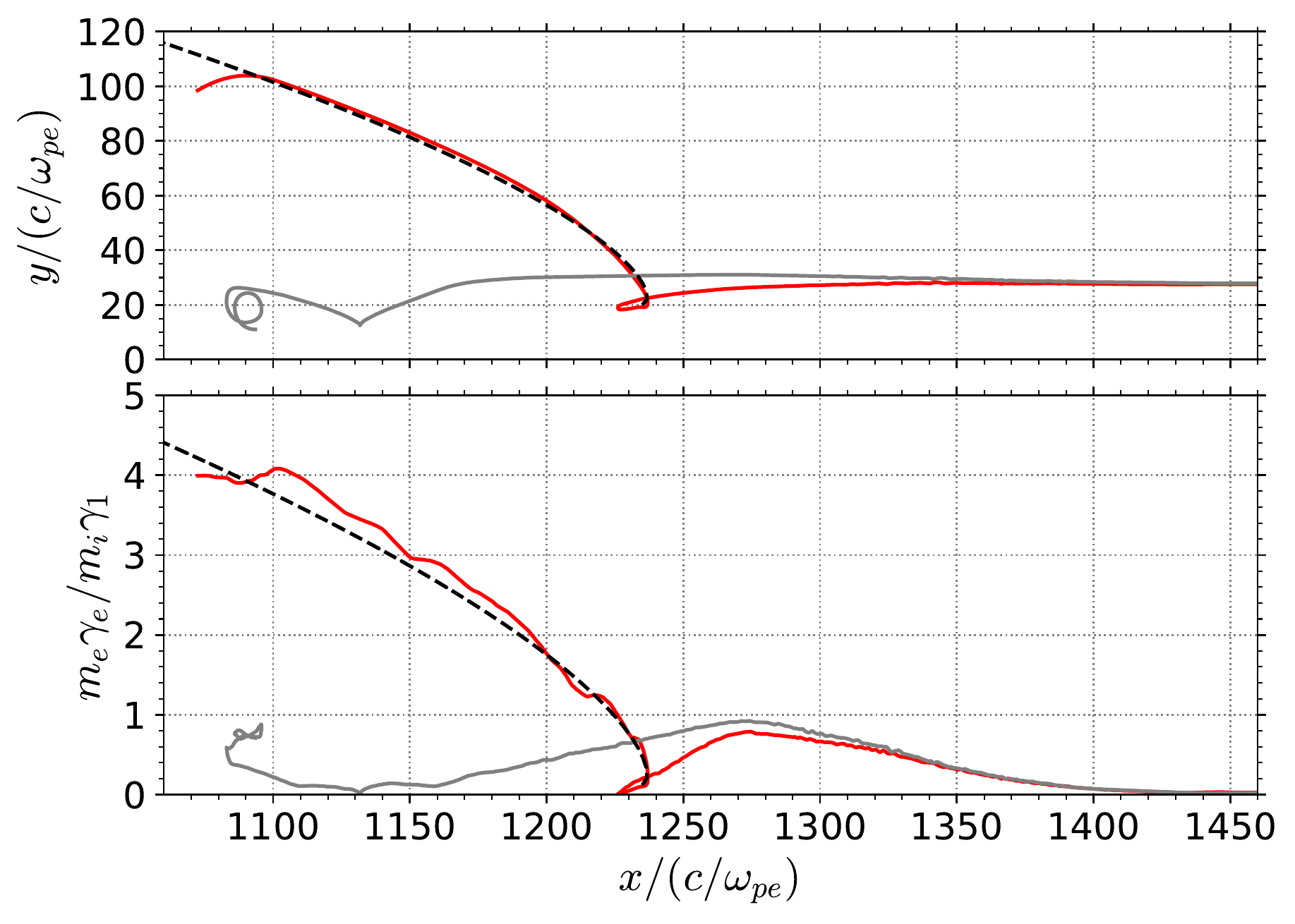}
 \caption{Nonthermal (red) and thermal (gray) electron trajectories in the
 $x$--$y$ (top) and $x$--$\gamma$ (bottom) space. The black dashed line is
 the analytical solutions of the pickup process.}
 \label{fig:orb_e}
\end{figure}

Here we discuss the electron acceleration mechanism.
Figure \ref{fig:orb_e} shows electron trajectories
measured in the simulation frame for $\gamma_1=40$
in the $x$--$y$ (top) and $x$--$\gamma$ (bottom)
space. Typical nonthermal and thermal
electrons are shown in the red and gray lines, respectively.
The thermal electron is merely oscillating within the wakefield.
It finally enters the shock at $x/(c/\omega_{pe}) \sim 1080$
and gyrates in the downstream region. The bulk Lorentz factor of the
thermal electrons has its maximum when the energy equipartition between
ions and electrons is achieved \citep{Lyubarsky2006,Hoshino2008,Iwamoto2019}
and thus it is written as
\begin{equation}
  \gamma_{max,e}^{th} \sim \frac{1}{2}\frac{m_i}{m_e}\gamma_1.
\end{equation}
As can be seen in the gray line in Figure \ref{fig:orb_e}, the normalized
maximum Lorentz factor is $m_e\gamma_e/m_i\gamma_1 \sim 1$. The difference
is at most a factor of 2 and
this result is roughly consistent with the above estimate.
However, the maximum Lorentz factor of the nonthermal electron is
much larger and can not be explained by the ion-electron coupling.
Although the nonthermal electron is
also oscillating in the region:
$1230 \lesssim x/(c/\omega_{pe}) \lesssim 1400$,
it begins to travel along the $+y$ direction at
$x/(c/\omega_{pe}) \sim 1230$ and seems to gain the energy from
the motional electric field $E_y = -\beta_1B_1$. This particle acceleration
continues until the electron enters the shock at
$x/(c/\omega_{pe}) \sim 1080$. Its maximum Lorentz factor $\gamma_{max,e}$
can be calculated as
\begin{equation}
  \gamma_{max,e} = \frac{e\beta_1B_1\Delta y}{m_ec^2}
  =\gamma_1\beta_1\sqrt{\sigma_e}\frac{\Delta y}{c/\omega_{pe}}.
\end{equation}
We measured $\Delta y/(c/\omega_{pe}) \sim 85$  in the region:
$1230 \lesssim x/(c/\omega_{pe}) \lesssim 1400$, and thus
$m_e\gamma_{max,e}/m_i\gamma_1 \sim 4$,
showing a good agreement with our simulation result.
This acceleration process is identical to the pickup process in space
physics \cite[e.g.,][]{Mobius1985,Oka2002}.
In the heliosphere, some neutrals are ionized mainly through charge
exchange with solar wind protons and picked up by the solar wind electric
field. The pickup ions are efficiently accelerated by the motional electric
field. The pickup process including the relativistic effect
can be theoretically analyzed using the relativistic equations of
motion. We have neglected the precursor waves,
wakefields, and filaments and
considered only the ambient magnetic field $B_z = B_1$ and the motional
electric field $E_y = -\beta_1B_1$. The analytical solution in the
simulation frame is written as
(see Appendix \ref{app:pickup} for the detailed calculations)
\begin{eqnarray}
  \label{eq:gam}
  \gamma_{s} &=& \gamma_1^2\gamma_{0s}\left[(1+\beta_1\beta_{0s})-
  \beta_1(\beta_1+\beta_{0s})\cos\theta_s  \right], \\
  \label{eq:x}
  x_s &=& x_{0s} -c\beta_1t+
  \gamma_{0s}(\beta_1+\beta_{0s})\frac{c}{\omega_{cs}}\sin\theta_s, \\
  \label{eq:y}
  y_s &=& y_{0s} \pm
  \gamma_1\gamma_{0s}(\beta_1+\beta_{0s})\frac{c}{\omega_{cs}}(1-\cos\theta_s), \\
  \label{eq:theta}
  \theta_s &=& \frac{\omega_{cs}[t+\beta_1(x_s-x_{0s})/c]}{\gamma_{0s}(1+\beta_1\beta_{0s})},
\end{eqnarray}
where the subscript $0$ indicates the initial quantities at the time
when the particles are picked up by the upstream bulk flow.
The positive (negative) sign in Equation \ref{eq:y} corresponds to electrons
(ions). Here we assume
$\bvec{\beta_{0s}} = +\beta_{0s}\bvec{\hat x}$.
Note that the upstream bulk flow propagates toward the $-x$ direction and
$\omega_{cs}$ is the unsigned cyclotron frequency.
We determined the initial quantities:
$x_{0s}$, $y_{0s}$, and $\beta_{0s}$ from our simulations and the
theoretical solutions are shown in Figure \ref{fig:orb_e}
by the black dashed lines. The simulation results give a good agreement
with the theoretical trajectories for the region:
$1080 \lesssim x/(c/\omega_{pe}) \lesssim 1240$, indicating that the
nonthermal electron is picked up by the bulk flow.
The electron reaches the shock front at $x/(c/\omega_{pe}) \sim 1080$
and gradually deviates from the theoretical trajectories.
According to Equation \ref{eq:gam}, the maximum Lorentz factor is estimated as
\begin{equation}
  \label{eq:gammax}
  \gamma_{max,s} \sim \gamma_1^2\gamma_{0s}.
\end{equation}
Here we have neglected the factors on the order of unity.

\begin{figure}[htb!]
 \plotone{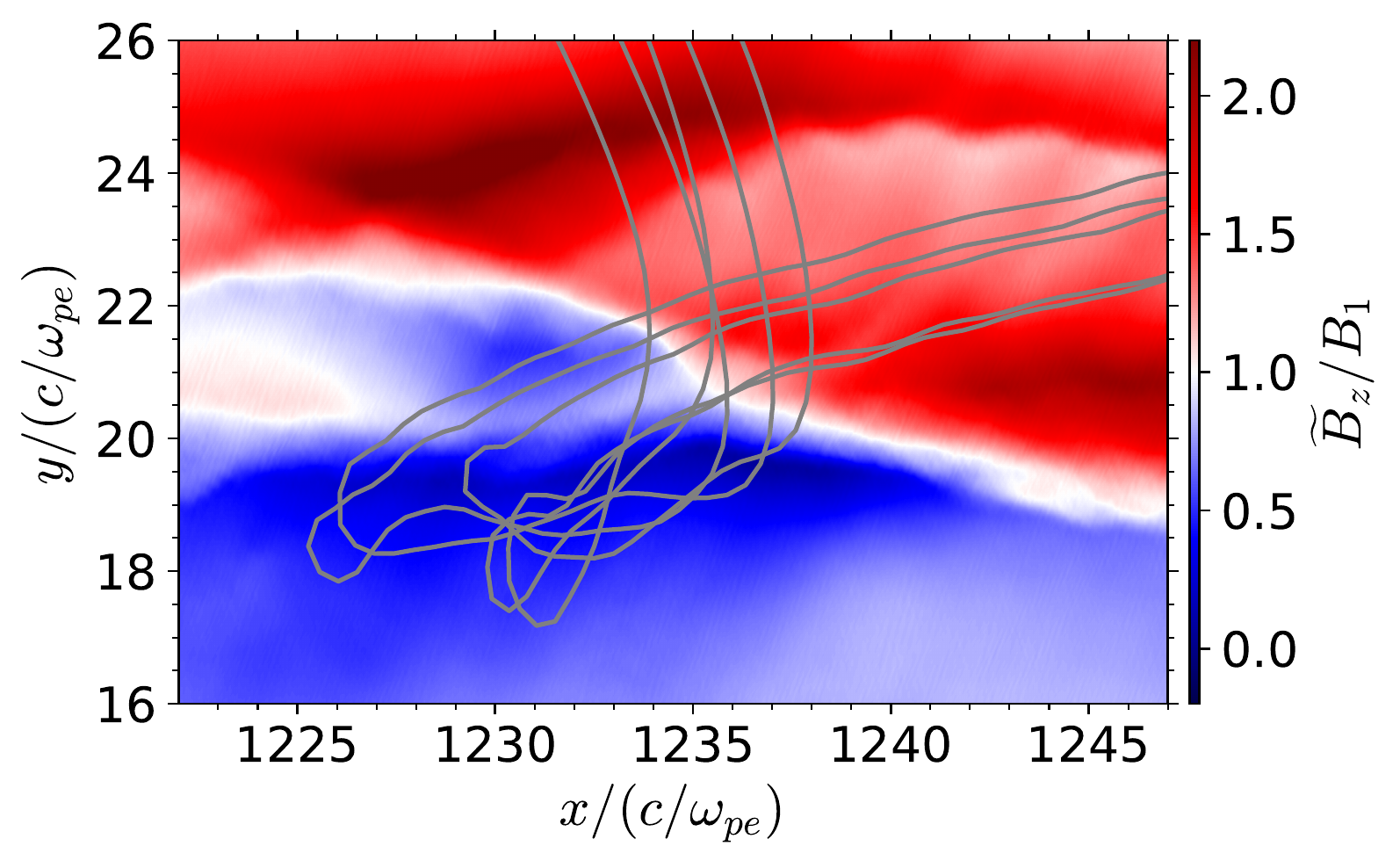}
 \caption{Trajectories of the nonthermal electrons with
 $\widetilde{B}_z$ at $\omega_{pe}t = 1800$.}
 \label{fig:bz_e}
\end{figure}

The above analytical solution indicates that the nonthermal particles
must travel in the opposite direction of the bulk flow before entering into
the pickup process. \cite{Sironi2011} reported the same acceleration process
and pointed out that such particles are decoupled from the bulk flow and
thus they can feel the motional electric field due to the velocity
difference. Equation \ref{eq:gam} indeed demonstrates that the particle
acceleration does not occur ($\gamma_s \sim \gamma_1$) for the particle
moving with the same velocity as the bulk flow $\beta_{0s} \sim -\beta_1$.
Figure \ref{fig:bz_e} shows the trajectories of nonthermal electrons
measured in the simulation frame in the gray lines. The color map
represents the magnetic field $\tilde{B}_z$ at $\omega_{pe}t=1800$, which
satisfies
\begin{equation}
  \tilde{E}_y = -\beta_1B_1- \frac{\partial \phi}{\partial y},
\end{equation}
\begin{equation}
  \tilde{E}_y + \beta_1\tilde{B}_z = 0.
\end{equation}
The electrostatic potential $\phi$ is calculated from the snapshot at
$\omega_{pe}t=1800$ by performing the Helmholtz's decomposition.
We have removed the electromagnetic fields arising from the
precursor waves because such superluminal waves are not responsible for
the resonant wave-particle interaction and do not directly
contribute to the particle acceleration.
The trajectories in Figure \ref{fig:bz_e} indeed demonstrate that
nonthermal electrons propagate toward the $+x$ direction before picked up by
the bulk flow. The nonthermal electrons come in to the weakly
magnetized region resulting from the FI and then enters the acceleration
phase, indicating that the filaments trigger the pickup process.

\begin{figure}[htb!]
 \plotone{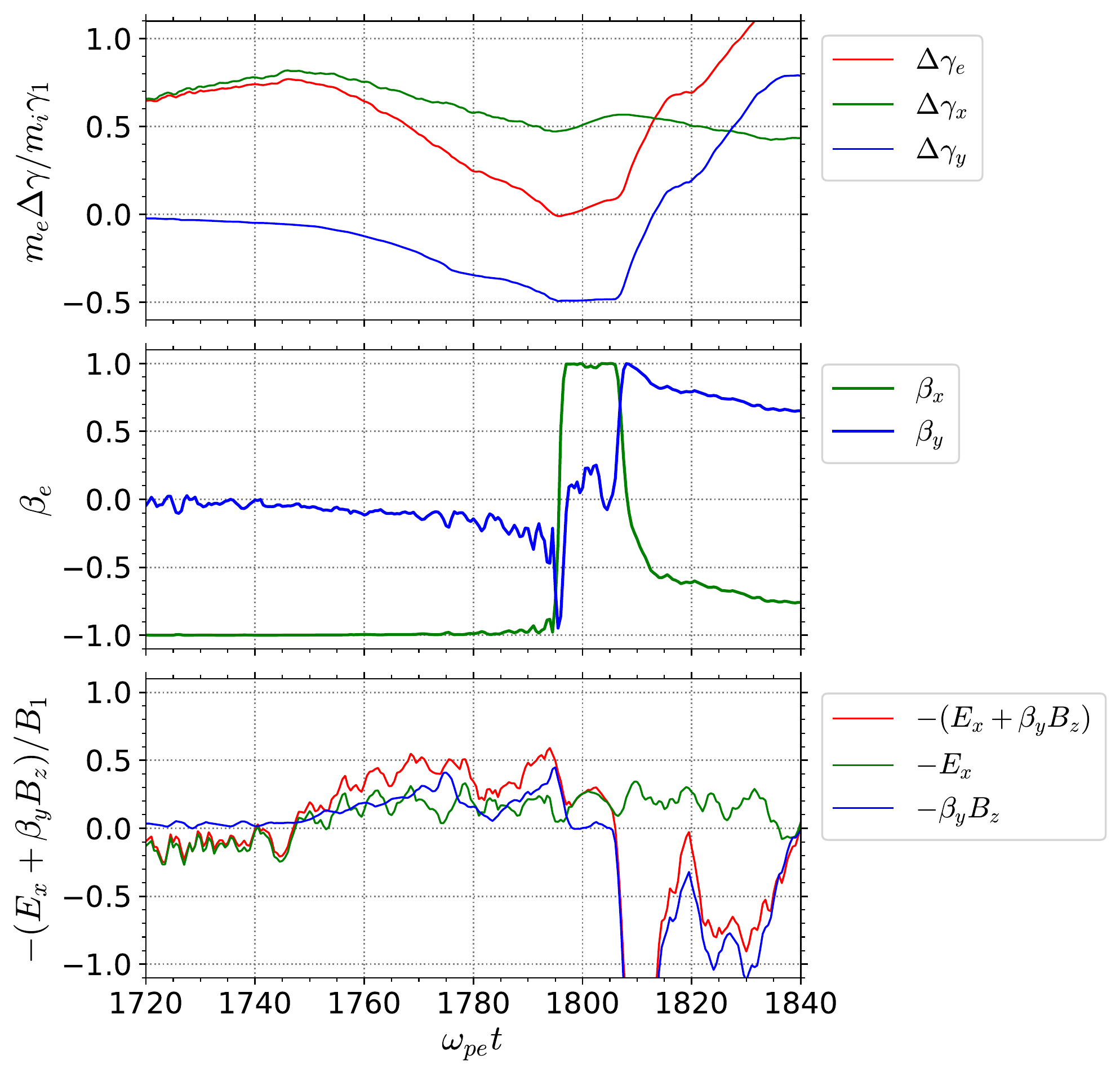}
 \caption{Time evolution of the nonthermal electron.
 Top panel: the energy gain $\Delta \gamma_e = \gamma_e-\gamma_1$ (red)
 and work done by $E_x$ (green) and $E_y$ (blue).
 Middle panel: the three velocity $\beta_x$ (green) and $\beta_y$ (blue).
 Bottom panel: the total force $-e(E_x+\beta_yB_z)$ (red),
 electrostatic force $-eE_x$ (green), and Lorentz force $-e\beta_yB_z$ (blue)
 at the electron position.}
 \label{fig:evo_e}
\end{figure}

Figure \ref{fig:evo_e} shows the time evolution of the typical
nonthermal electron measured in the simulation frame. We take the moving
average for the time period $\omega_{pe}\Delta t =5$, which is
motivated by the typical frequency of the precursor waves:
$\omega/\omega_{pe} \sim 2-5$ \citep{Iwamoto2017,Iwamoto2018},
to remove the effect
of the precursor waves.
The top panel displays the energy gain
$\Delta \gamma_e = \gamma_e-\gamma_1$ (red) and
work done by $E_x$ (green) and $E_y$ (blue) normalized by $m_i\gamma_1/m_e$:
\begin{equation}
  \Delta \gamma_k = -\frac{e}{m_ec^2}\int^t_{t_0}E_kv_k{\rm d}t,
\end{equation}
where $k=x,y$ and $\omega_{pe}t_0=1500$.
The incoming electron is decelerated by the wakefield and begin to gyrate.
Then it loses its energy due to both the wakefield and the motional
electric field for $1750 \lesssim \omega_{pe}t \lesssim 1795$.
$\Delta \gamma_x$ increases in time for
$1795 \lesssim \omega_{pe}t \lesssim 1810$, wheres $\Delta \gamma_y$ is almost
constant, showing that the energy gain during the corresponding time period
originates from the wakefield. After $\omega_{pe}t \simeq 1810$,
$\Delta \gamma_y$ becomes dominant and thus the nonthermal electron enters the
pickup process. The $x$ and $y$ components of the electron velocity
normalized by the speed of light:
$\beta_x$ (green) and $\beta_y$ (blue) are shown in the middle panel.
$\beta_x$ is positive for $1795 \lesssim \omega_{pe}t \lesssim 1810$ and
the nonthermal electron moves with the relativistic velocity
$\beta_x \sim 1$ in the same direction of the wakefield propagation. Note
that the phase velocity of the wakefield is almost equal to the speed of
light \citep{Hoshino2008}. $B_z \sim 0$ is satisfied inside the filaments
and the electrostatic force $-eE_x$ easily overcomes the Lorentz force
$-e\beta_yB_z$ despite $E_{wake}/B_1 < 1$.
Therefore, the electron is trapped by the wakefield and accelerated
via the Landau resonance.
The acceleration
continues until the Lorentz force exceeds the electrostatic force.
The bottom panel of Figure \ref{fig:evo_e} shows the total force
$-e(E_x+\beta_yB_z)$ (red), electrostatic force $-eE_x$ (green), and
Lorentz force $-e\beta_yB_z$ (blue) normalized by $eB_1$ at the electron
position. The electrostatic force indeed dominates over the Lorentz force
for $1795 \lesssim \omega_{pe}t \lesssim 1810$ and then the total force is
controlled by the Lorentz force after $\omega_{pe}t \simeq 1810$. One may
think that this process is similar to the shock surfing acceleration
\cite[SSA;][]{Shimada2000,Hoshino2002}, in
which the electrostatic waves trap the incoming electrons and the motional
electric field accelerates them during the multiple reflection within the
electrostatic waves. The essential difference is that the electrons gain
the energies from the wakefield during the trapping because the motional
electric field vanishes within the filaments $B_z \sim 0$.
This acceleration mechanism is analogous to the standard WFA in laboratory
plasmas rather than the SSA. We have confirmed our idea that the electrons
pre-accelerated by the wakefield inside the filaments are further
accelerated via the pickup process by performing the test particle
simulations (see Appendix \ref{app:test}).

We evaluate the initial Lorentz factor $\gamma_{0e}$.
Since the wakefield directly accelerates the nonthermal electron within the
filaments, the energy gain is expressed as
\begin{equation}
 \label{eq:dgam}
 \Delta \gamma_x =
 \frac{eE_{wake}L_{acc,e}}{m_ec^2}
 = \gamma_1\sqrt{\sigma_e}\frac{E_{wake}}{B_1}\frac{L_{acc,e}}{c/\omega_{pe}},
\end{equation}
where $E_{wake}$ is the wakefield amplitude and $L_{acc,e}$ is the acceleration
length. $E_{wake}$ can be estimated as
\citep{Hoshino2008}
\begin{equation}
  \label{eq:wake}
  \frac{E_{wake}}{B_1} = \frac{1}{\gamma_1\sqrt{\sigma_e}}
  \frac{\eta a^2}{\sqrt{1+\eta a^2}} \sim
  \sqrt{\frac{\epsilon_p}{\sigma_e}},
\end{equation}
where $\eta$ represents the wave polarization: $\eta=1$ for circular
polarization and $\eta=1/2$ for linear polarization.
Here we have used Equation \ref{eq:a} and $\eta=1/2$, and
neglected factors on the order of unity. By substituting ${\epsilon_p \sim 1}$
and $\sigma_e = 5$ into Equation \ref{eq:wake}, we obtain $E_{wake}/B_1 \sim O(10^{-1})$,
which agrees with our simulation results (see Figure \ref{fig:shock}).
The acceleration length corresponds to the wakefield wavelength for the WFA in
laboratory plasmas. In our shock simulations, however, the acceleration length
is limited by the size of the unmagnetized region
which is much smaller than the wakefield wavelength.
We evaluate $L_{acc,e}$ from the nonthermal electron trajectories in
Figure \ref{fig:bz_e}:
\begin{equation}
  \label{eq:lacc}
  \frac{L_{acc,e}}{c/\omega_{pe}} \sim 10.
\end{equation}
By substituting Equations \ref{eq:wake} and \ref{eq:lacc} into Equation \ref{eq:dgam},
we have the estimate of the initial Lorentz factor $\gamma_{0e}$:
\begin{equation}
  \gamma_{0e} \sim 1+\Delta \gamma_x \sim \alpha\gamma_1\sqrt{\epsilon_p},
\end{equation}
where
\begin{equation}
  \alpha \equiv \frac{L_{acc,e}}{c/\omega_{pe}} \sim 10.
\end{equation}
By substituting $\epsilon_p \sim 1$, we have
$m_e\Delta \gamma_x/m_i\gamma_1 \sim \alpha\sqrt{\epsilon_p}m_e/m_i \sim O(10^{-1})$.
The increase of $\Delta \gamma_x$ for $1795 \lesssim \omega_{pe}t \lesssim 1810$
is $\sim 0.1$ (see the top panel of Figure \ref{fig:evo_e})
and consistent with this estimate. We finally obtain
\begin{equation}
  \label{eq:gammaxe}
  \gamma_{max,e} \sim \alpha\gamma_1^3\sqrt{\epsilon_p} \sim \alpha\gamma_1^3.
\end{equation}
This shows that a highly relativistic shock $\gamma_1 \gg 1$ can be
an efficient particle accelerator.

Equation \ref{eq:gammaxe} shows the electron maximum Lorentz factor is
$ \gamma_{max,e}m_e/m_i \sim 10^4$ for $\gamma_1 = 40$ and
$ \gamma_{max,e}m_e/m_i \sim 10^6$ for $\gamma_1 = 100$.
However, the electron energy spectra in Figure \ref{fig:spectra}
demonstrate that
the maximum Lorentz factor in our simulations
is much smaller than we expect.
Although the maximum Lorentz factor for $\gamma_1 = 100$ is larger
than that for $\gamma_1 = 40$, the difference is at most a factor of 2.
The deviation from the analytical estimate can be explained as follows.
In the simulation frame, the picked-up electrons propagate towards the $-x$
direction while accelerated by the motional electric
field. If they are picked up near the shock front,
they enter the shock soon and the acceleration ceases before they take the
maximum Lorentz factor.
Equation \ref{eq:gam} shows that the Lorentz factor takes its maximum when
$\theta_e=\pi$. Equation \ref{eq:x} reduces to
\begin{equation}
  \label{eq:xe}
  x_e = x_{0e} -c\beta_1t_{acc,e},
\end{equation}
where $t_{acc,e}$ is the acceleration timescale.
By substituting this and $\theta_e=\pi$ into Equation \ref{eq:theta},
we can evaluate $t_{acc,e}$:
\begin{equation}
  \omega_{ce}t_{acc,e} \sim 2\pi \gamma_1^2\gamma_{0e}
  \sim 2 \pi \alpha \gamma_1^3,
\end{equation}
We finally obtain the moving distance of the electron in the $x$ direction
$\Delta x_e = |x_e-x_{0e}|$ during the time period $\Delta t=t_{acc,e}$:
\begin{equation}
  \frac{\Delta x_e}{c/\omega_{pe}} = \beta_1\omega_{pe}t_{acc,e}
  \sim \frac{2 \pi \alpha \gamma_1^3}{\sqrt{\sigma_e}}.
\end{equation}
This estimate shows $\Delta x_e/(c/\omega_{pe}) \sim 10^6$ for
$\gamma_1=40$ and $\Delta x_e/(c/\omega_{pe}) \sim 10^7$ for $\gamma_1=100$,
which are much larger than the precursor wave region:
$1100 \lesssim x/(c/\omega_{pe}) \lesssim 1800$
in the final state of our simulations.
Therefore, the picked-up electrons enter the shock before they obtain the
theoretical maximum Lorentz factor. Since the group velocity of the
precursor wave $v_{g} \sim c$ is faster than the shock propagation velocity
$v_{sh} \sim (1/2+3\sigma_i/4)c \sim 0.575c$, the precursor wave region
becomes larger as time passes. In the later phase,
the incoming electrons can be picked up far away from the shock front and
sufficiently accelerated by the motional electric field before entering the
shock. We thus think that the observed Lorentz factor will be closer to
the theoretical one if we follow the long-term evolution. Our test-particle
simulations indeed demonstrate that the maximum Lorentz factor
is consistent with the above estimate (see Appendix \ref{app:test}).
These results indicate that the particle energy spectra do not reach the
steady state yet. Nonthermal tails might be observed in the
downstream in the later phase.

\subsection{Ion acceleration}\label{subsec:i}

\begin{figure}[htb!]
 \plotone{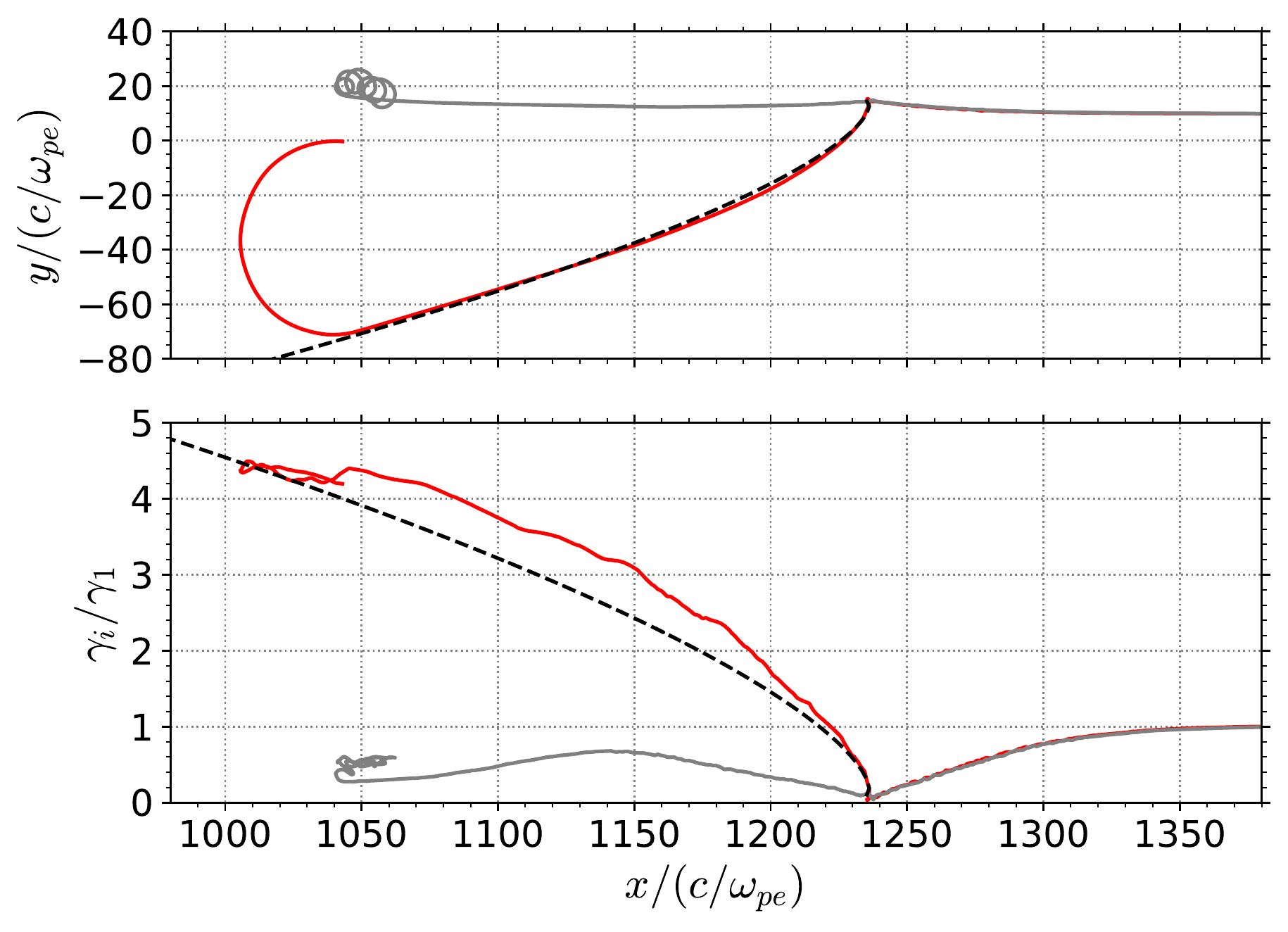}
 \caption{Ion trajectories in the same format as Figure \ref{fig:orb_e}}
 \label{fig:orb_i}
\end{figure}

The ion acceleration can be explained by the pickup process as well.
Figure \ref{fig:orb_i} shows the nonthermal and thermal ion trajectories
in the same format as Figure \ref{fig:orb_e}. The thermal ion is
oscillating inside the wakefield, whereas the nonthermal ion is
accelerated by the motional electric field.
The analytical solutions of the pickup process (Equations \ref{eq:gam},
\ref{eq:x}, \ref{eq:y}, and \ref{eq:theta}) are shown in black and gives a
good agreement with the simulation results.

\begin{figure}[htb!]
 \plotone{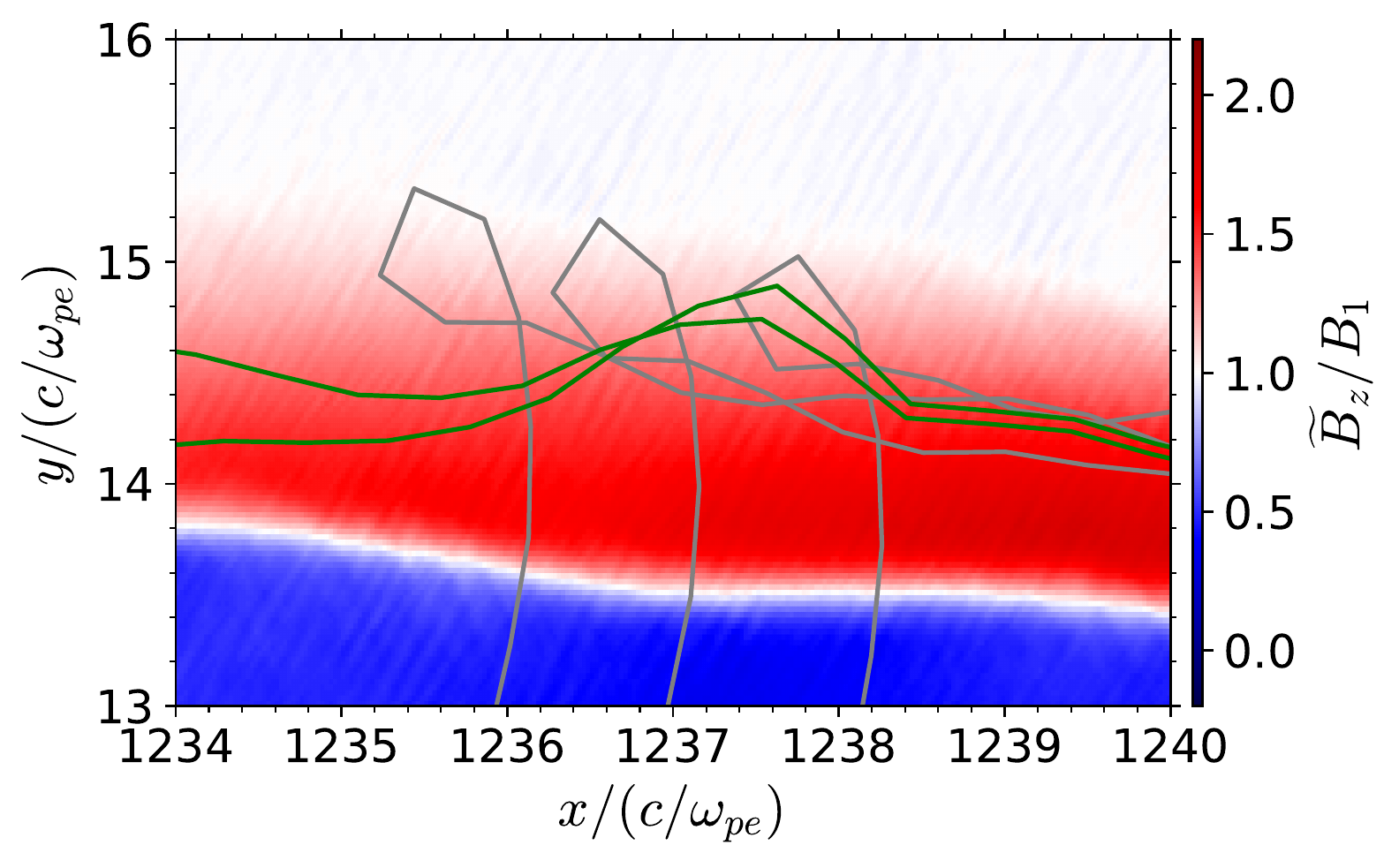}
 \caption{Trajectories of the nonthermal (gray) and thermal (green)
 ions with $\widetilde{B}_z$ at $\omega_{pe}t = 1660$.}
 \label{fig:bz_i}
\end{figure}

We here discuss how ions are injected into the pickup process.
Figure \ref{fig:bz_i} shows the trajectories of nonthermal (gray) and
thermal (green) ions with the magnetic field $\tilde{B}_z$ which is
determined from the snapshot at $\omega_{pe}t = 1660$ in the same manner as
for the electron. The ion injection occurs in highly magnetized region
$B_z/B_1>1$ unlike the electron. The incoming cold ions are gradually
thermalized by the SRS and/or the FI, which is clearly seen in the phase
space density plots of Figure \ref{fig:shock}. The thermalized ones
can be slightly deviated from the bulk motion and $E \times B$ drift can be
induced. The ion's trajectory is thus given by the cycloid, which is the
case for the thermal ions (green lines).
On the other hand, the nonthermal ones (gray lines) are suddenly
reflected toward the $+x$ direction during the cycloid motion and then
picked up by the bulk flow. This kick toward upstream
seems to trigger the pickup process.

\begin{figure*}[htb!]
 \plottwo{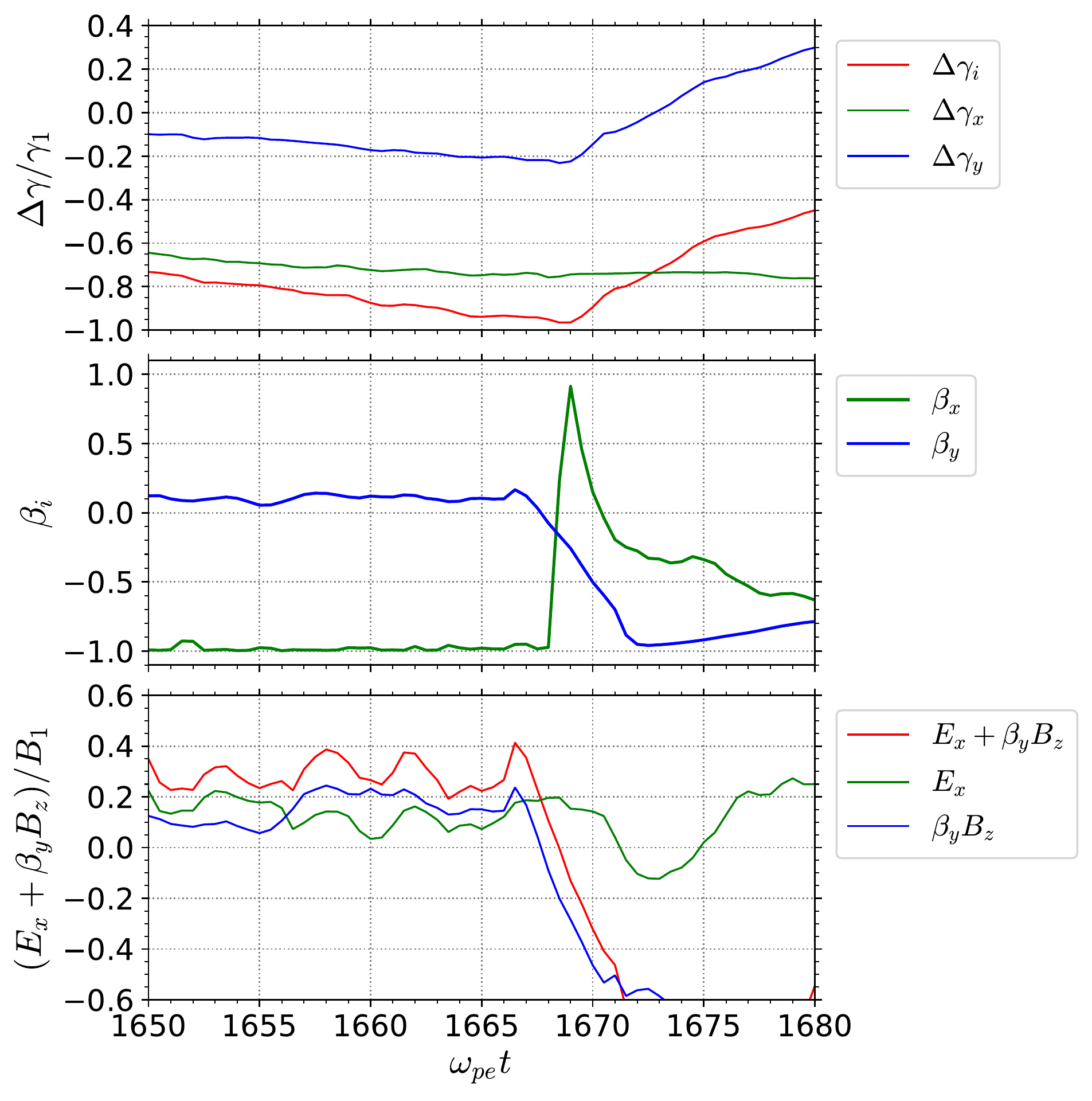}{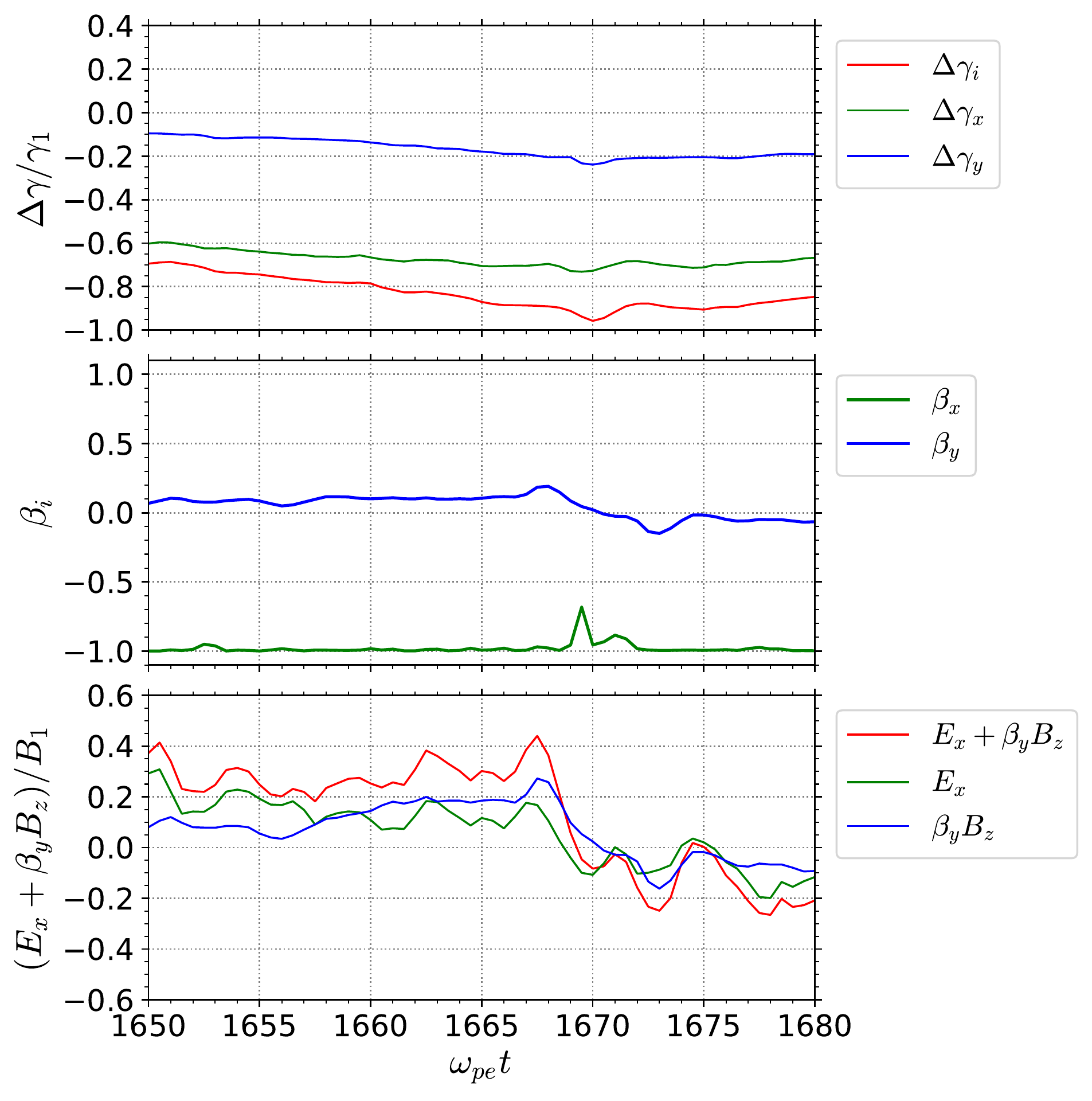}
 \caption{Time evolution of the nonthermal (left) and thermal (right)
 ion in the same format as Figure \ref{fig:evo_e}}
 \label{fig:evo_i}
\end{figure*}

Figure \ref{fig:evo_i} displays the time evolution of the typical
nonthermal (left) and thermal (right) ions in the same format as Figure
\ref{fig:evo_e}. We take the moving average for the time period
$\omega_{pe}\Delta t =5$ as well. The energy gain
$\Delta \gamma_i = \gamma_i-\gamma_1$ (red) and
work done by $E_x$ (green) and $E_y$ (blue) normalized by $\gamma_1$
are shown in the top panels.
In the case of the nonthermal ions (left),
both the wakefield and the motional electric field contribute to the energy
loss for $\omega_{pe}t \lesssim 1669$.
$\Delta \gamma_i$ increases in time after $\omega_{pe}t \sim 1669$ and
$\Delta \gamma_y$ exhibits the same tendency,
indicating that the ion enters the pickup process at
$\omega_{pe}t \sim 1669$. As can be seen in the left middle panel,
at $\omega_{pe}t \sim 1669$, $\beta_x$ becomes positive and the ion
decoupling occurs. The left bottom panel of Figure \ref{fig:evo_e}
demonstrates that the electrostatic force $eE_x$ (green) exceeds the
Lorentz force $e\beta_yB_z$ (blue) for
$1668 \lesssim \omega_{pe}t \lesssim 1669$ and
the wakefield can reflect the incoming ion.
We think that the kick imparted by the wakefield
determines whether the ion enters the
pickup process or not. The drifting ions can satisfy
$\beta_y \sim 0$ at some point on the way to the shock.
If the wakefield kicks them at the time when $\beta_y \sim 0$
is satisfied, the electrostatic force $eE_x$ can easily overcome
the Lorentz force $e\beta_yB_z \sim 0$.
Furthermore, the Lorentz factor of the drifting ions has the minimum value
when $\beta_y \sim 0$ and they are relatively subject to the wakefield.
In fact, the thermal ion (right) shows that the $eE_x$ is negative
at the time $\omega_{pe}t \sim 1670$ when $\beta_y \sim 0$ is
satisfied and the wakefield cannot reflect it.
Although the increase of $\Delta \gamma_x$ is barely visible for
$1668 \lesssim \omega_{pe}t \lesssim 1669$ in the left top panel,
we think the finite
kick imparted by the wakefield is responsible for the decoupling.

The ion injection into the pickup process seems to be different
from the electron. This may be attributed to the mass
difference like the standard WFA in laboratory plasmas. Electrons are
relatively easily accelerated by the wakefield whose phase velocity is
almost equal to the speed of light via the Landau resonance, whereas ions
have difficulty with the resonance due to the large mass.
Our simulations indeed demonstrate the efficient electron WFA inside the
filaments. Although the ion WFA is transient and
inefficient, we think that this finite pre-acceleration injects
ions into the pickup process.

We here evaluate the ion maximum Lorentz factor $\gamma_{max,i}$ in the same manner as for
the electron. Since the incoming ion is reflected by the wakefield,
$\Delta \gamma_x$ can be written as
\begin{equation}
  \Delta \gamma_x = \frac{eE_{wake}L_{acc,i}}{m_ic^2}
  \sim \frac{m_e}{m_i}\gamma_1\sqrt{\epsilon_p}\frac{L_{acc,i}}{c/\omega_{pe}}.
\end{equation}
The nonthermal ion trajectories in Figure \ref{fig:bz_i} indicate
\begin{equation}
  \frac{L_{acc,i}}{c/\omega_{pe}} \sim 1.
\end{equation}
$\gamma_{0i}$ can be estimated as
\begin{equation}
  \gamma_{0i} \sim 1+\Delta \gamma_x \sim 1+\frac{m_e}{m_i}\gamma_1\sqrt{\epsilon_p}
\end{equation}
The ion maximum Lorentz factor $\gamma_{max,i}$ can be derived from Equation
\ref{eq:gammax},
\begin{equation}
  \label{eq:gammaxi}
  \gamma_{max,i}
  \sim \left(1+\frac{m_e}{m_i}\gamma_1\sqrt{\epsilon_p}\right)\gamma_1^2
  \sim \left(1+\frac{m_e}{m_i}\gamma_1\right)\gamma_1^2
\end{equation}
Here we have used $\epsilon_p \sim 1$.
As can be seen in Figure \ref{fig:spectra},
the ion energy spectra show the smaller maximum Lorentz factor due to the
time dilation.
We can estimate the moving distance in the $x$ direction
during the acceleration as in the case of the electron.
The accceleration timescale $t_{acc,i}$ is expressed as
\begin{equation}
  \omega_{ci}t_{acc,i} \sim
  2\pi \left(1+\frac{m_e}{m_i}\gamma_1\right)\gamma_1^2.
\end{equation}
We obtain the moving distance of the ion in the $x$ direction $\Delta x_i = |x_i-x_{0i}|$:
\begin{equation}
  \frac{\Delta x_i}{c/\omega_{pe}} = \beta_1\omega_{pe}t_{acc,i}
  \sim \frac{2\pi \left(m_i/m_e+\gamma_1\right)\gamma_1^2}{\sqrt{\sigma_e}}.
\end{equation}
This estimate gives
$\Delta x_i/(c/\omega_{pe})\sim 10^5$ for
$\gamma_1 = 40$ and $\sim 10^6$ for $\gamma_1 = 100$ and thus
the pickup ions enter the shock before they reach the maximum Lorentz
factor in our simulations. Nonthermal ions as well as electrons might be
seen in the downstream in the later phase.

\section{Discussion}\label{sec:discuss}

In this work, we assumed the precursor wave power $\epsilon_p \sim 1$ which
is valid for $\sigma_i \sim 0.1 - 1$ \citep{Iwamoto2019}.
$\epsilon_p$ is independent of $\gamma_1$ as long as $\gamma_1 \gg 1$
\citep{Plotnikov2019} and it is mainly controlled by $\sigma_i$ due to the
ion-electron coupling \citep{Lyubarsky2006,Hoshino2008,Iwamoto2019}.
Although the $\sigma_i$ dependence is not fully understood, previous PIC
simulations demonstrated that $\epsilon_p$ is convex upward as a function
of $\sigma_i$ and takes the maximum value $\epsilon_p \sim 1$ at
$\sigma_i \sim 0.1$.
The ion acceleration efficiency (Equation \ref{eq:gammaxi}) is not
strongly dependent on $\epsilon_p$ as long as $\gamma_1 < m_i/m_e$ and
$\gamma_{max,i} \sim \gamma_1^2$ for $\epsilon_p \ll 1$.
On the other hand, the electron acceleration efficiency (Equation
\ref{eq:gammaxe}) drastically deteriorates for $\epsilon_p \ll 1$
and may be reduced to $\gamma_{max,e} \sim \gamma_{max,i} \sim \gamma_1^2$.
For low $\sigma_i$, however, the electron acceleration is not necessarily
less efficient. The acceleration length $L_{acc,e}$
may be much greater because $E_{wake}/B_1 \sim \delta B/B_1>1$ can be
satisfied for weakly magnetized plasmas and
the electrostatic force can easily exceed the Lorentz force.
The acceleration length may be comparable to
the wakefield wavelength \citep{Kruer1988,Hoshino2008},
\begin{equation}
  \alpha = \frac{L_{acc,e}}{c/\omega_{pe}} \sim \gamma_1.
\end{equation}
The maximum Lorentz factor is expressed as
\begin{equation}
  \label{eq:gam4}
  \gamma_{max,e} \sim \gamma_1^4\sqrt{\epsilon_p}.
\end{equation}
The Weibel instability develops for $\sigma_i \ll 1$ and the wave power
declines because the ring-like momentum
distribution in the shock-transition which is essential for the SMI is
strongly modified by the Weibel-generated magnetic field
\citep{Sironi2011,Iwamoto2017,Iwamoto2018}.
However, Equation \ref{eq:gam4} exhibits the weak dependence on $\epsilon_p$
compared to $\gamma_1$.
We thus speculate that the efficient electron acceleration
occurs as long as $\gamma_1 \gg 1$.

The upstream temperature has an influence on the acceleration efficiency as
well. Since the accelerated/heated particles enter the shock, $\epsilon_p$
may decrease in time due to the suppression of the higher-order harmonics
\citep{Amato2006}. In pair plasma, \cite{Babul2020} reported that
the wave emission efficiency declines by almost two orders of magnitude for
the thermal spread $k_BT_e/m_ec^2 \gtrsim 10^{-1}$.
Although the temperature dependence in ion-electron plasmas remains
unsolved, $\epsilon_p$ probably shows the similar tendency.
The precursor wave emission might cease and the size of the
precursor wave region might be insufficient to accelerate the incoming
particles up to the theoretical estimate even if we follow the long-term
evolution. The particle energy spectra in the final state is an open
question.

Both electrons and ions are accelerated via the pickup process
in the upstream. We speculate that the pickup process provides seed
particles for other acceleration mechanisms such as Fermi acceleration.
The pre-accelerated particles in the upstream may be further accelerated
and power-low spectra may be generated in the downstream.

The pre-existing cosmic rays may be re-accelerated by the pickup process.
Such energetic protons can diffuse far upstream. Since they are
decoupled from the upstream bulk flow, the pickup process can work.
According to Equation \ref{eq:gammax}, they can be re-accelerated by a
factor of $\gamma_1^2$. This re-acceleration process may repeatedly
operate and they may be accelerated up to the UHECR energy range.

\section{Summary}\label{sec:summary}

We investigated the particle acceleration in relativistic
ion-electron shocks by 2D PIC simulations.
The particle energy spectra in the upstream show the nonthermal tails
for both electrons and ions.
We found that they are mainly accelerated by the motional electric field.
This particle acceleration is well-described by the pickup process,
in which particles are once decoupled from the upstream bulk flow by the
wakefield, and are piked up again by the flow.
We estimated the maximum Lorentz factor
$\gamma_{max,e} \sim \alpha\gamma_1^3$ for the electron and
$\gamma_{max,i} \sim (1+m_e\gamma_1/m_i)\gamma_1^2$ for the ion,
where $\alpha \sim 10$ is the normalized acceleration length and
determined from our simulations.
Since this acceleration requires a large computational domain,
we could not follow the whole acceleration process
due to the limitation of the computational resources.
The accelerated particles might exhibit a power-law-like spectra
in the downstream in a later phase.
The pickup process may play a significant role for particle acceleration
in highly relativistic shocks $\gamma_1 \gg 1$ such as external shocks of
GRBs.

\acknowledgments
We are grateful to Jacek Niemiec, Martin Pohl, Oleh Kobzar,
Arianna Ligorini, and Artem Bohdan for fruitful discussions.

This work is supported by JSPS KAKENHI grant No. 20J00280 and 20KK0064.

This work used the computational resources of the HPCI system provided by
Information Technology Center, Nagoya University through the HPCI System Research
Project (Project ID: hp200035, hp210154).

Numerical computations were in part carried out on Cray
XC50 at Center for Computational Astrophysics, National Astronomical Observatory
of Japan.

\appendix
\section{Analytical solutions of pickup process}\label{app:pickup}
We here derive the analytical solutions of the pickup process. Let us
assume a charged particle in the background magnetic field $B_z = B_1$ and
the motional electric field $E_y = -\beta_1B_1$.
The basic equations are the relativistic equations of motion:
\begin{eqnarray}
  \label{eq:ux}
  m_sc\frac{{\rm d}u_{xs}}{{\rm d}t} &=& q_s\beta_{ys}B_1, \\
  \label{eq:uy}
  m_sc\frac{{\rm d}u_{ys}}{{\rm d}t} &=& -q_s(\beta_{xs}+\beta_1)B_1.
\end{eqnarray}
where $\bvec{u_s}= \gamma_s\bvec{\beta_s}$ is the four velocity,
$q_e=-e$ is the electron charge, and $q_i = +e$ is the ion charge.
By performing Lorentz transformation from the simulation frame
into the plasma rest frame,
Equations \ref{eq:ux} and \ref{eq:uy} reduce to
\begin{eqnarray}
  \label{eq:uxp}
  m_sc\frac{{\rm d}u_{xs}^{\prime}}{{\rm d}t^{\prime}}
  &=& q_s\beta_{ys}^{\prime}B_1^{\prime} \\
  \label{eq:uyp}
  m_sc\frac{{\rm d}u_{ys}^{\prime}}{{\rm d}t^{\prime}}
  &=& -q_s\beta_{xs}^{\prime}B_1^{\prime},
\end{eqnarray}
where the prime indicates the physical quantities in the plasma rest frame.
The motional electric field vanishes in the plasma rest
frame and thus the kinetic energy is conserved:
\begin{equation}
  \gamma_s^{\prime}  = const. = \gamma_1\gamma_{0s}(1+\beta_1\beta_{0s}),
\end{equation}
where $\gamma_{0s} = 1/\sqrt{1-\beta_{0s}^2}$ is the initial Lorentz fator at
$t=0$. Here we have assumed the initial three velocity
$\bvec{\beta_{0s}} =\beta_{0s}\bvec{\hat x}$ and performed the Lorentz transformation
$\gamma_s^{\prime}  = \gamma_1\gamma_s(1+\beta_1\beta_{xs})$.
Equations \ref{eq:uxp} and \ref{eq:uyp} describe the gyromotion
around the background magnetic field $B_1^{\prime} = B_1/\gamma_1$
and are easily solved:
\begin{eqnarray}
  u_{xs}^{\prime} &=& u_{0s}^{\prime}\cos\theta_s^{\prime}, \\
  u_{ys}^{\prime} &=& \pm u_{0s}^{\prime}\sin\theta_s^{\prime} , \\
  x_{s}^{\prime} &=& x_{0s}^{\prime} +
  u_{0s}^{\prime}\frac{c}{\omega_{cs}}\sin\theta_s^{\prime}, \\
  y_{s}^{\prime} &=& y_{0s}^{\prime} \pm
  u_{0s}^{\prime}\frac{c}{\omega_{cs}}(1-\cos\theta_s^{\prime}), \\
  \theta_s^{\prime} &=& \frac{\omega_{cs}t^{\prime}}{\gamma_s^{\prime}},\\
  u_{0s}^{\prime} &=& \gamma_1\gamma_{0s}(\beta_1+\beta_{0s}),
\end{eqnarray}
where the subscript $0$ represents the initial quantities at $t=0$.
The positive (negative) sign corresponds to the electron (ion).
Note that $\omega_{cs} = eB_1/\gamma_1m_sc$ is the unsigned cyclotron
frequency.
By performing Lorentz transformation from the plasma rest frame
into the simulation frame,
we obtain the exact solutions of Equations \ref{eq:ux} and \ref{eq:uy}:
\begin{eqnarray}
  \label{eq:ga,2}
  \gamma_{s} &=& \gamma_1^2\gamma_{0s}\left[(1+\beta_1\beta_{0s})-
  \beta_1(\beta_1+\beta_{0s})\cos\theta_s  \right], \\
  \label{eq:ux1}
  u_{xs} &=& \gamma_1^2\gamma_{0s}\left[-\beta_1(1+\beta_1\beta_{0s})+
  (\beta_1+\beta_{0s})\cos\theta_s
  \right], \\
  \label{eq:uy2}
  u_{ys} &=& \pm \gamma_1\gamma_{0s}(\beta_1+\beta_{0s})\sin\theta_s, \\
  \label{eq:x2}
  x_s &=& x_{0s} -c\beta_1t+
  \gamma_{0s}(\beta_1+\beta_{0s})\frac{c}{\omega_{cs}}\sin\theta_s, \\
  \label{eq:y2}
  y_s &=& y_{0s} \pm
  \gamma_1\gamma_{0s}(\beta_1+\beta_{0s})\frac{c}{\omega_{cs}}(1-\cos\theta_s), \\
  \label{eq:theta2}
  \theta_s &=& \frac{\omega_{cs}[t+\beta_1(x_s-x_{0s})/c]}{\gamma_{0s}(1+\beta_1\beta_{0s})}.
\end{eqnarray}
By numerically solving Equations \ref{eq:x2} and \ref{eq:theta2},
we can determine $x_s$ and $\theta_s$ and finally obtain the theoretical solutions
of the pickup process.

\section{Test particle simulation}\label{app:test}

To confirm that the transverse filamentary structures
triggers the pickup process,  we perform the test particle simulations.
The particle pusher proposed by \cite{Vay2008} is applied to this
test
particle code. We consider the ambient magnetic field $B_1$, the motional
electric field
$E_y = -\beta_1B_1$, the wakefield, and the filaments.
The wakefield and filaments are modeled as
\begin{equation}
  E_x=\left\{
  \begin{array}{ll}
    -E_{wake}\sin\left[\frac{2\pi}{\lambda_{wake}}(x-ct)\right] & (x < ct) \\
    0 & (x \ge ct)
  \end{array}
  ,\right.
\end{equation}
\begin{eqnarray}
  B_z &=& B_1 + B_f\sin\left(\frac{2\pi}{\lambda_{f}}y\right), \\
  E_y &=& -\beta_1B_z,
\end{eqnarray}
where $E_{wake}$ is the wakefield amplitude, $\lambda_{wake}$ is the
wakefield wavelength,  $B_f$ is the filament amplitude, and $\lambda_f$ is the
filament wavelength. Particles are injected at $x=0$  toward the $-x$ direction
with the bulk  Lorentz factor $\gamma_1=40$.
The thermal velocity of the injected plasma flow in the plasma rest frame
is $\beta_{th} = 0.1$.
Based on our PIC simulation results,
we determined $E_{wake}/B_1 = 0.2$, $\lambda_{wake}/(c/\omega_{pe}) = 500$,
$B_f/B_1 = 1$, and $\lambda_f/(c/\omega_{pe}) = 15$.
The other parameters are identical to our PIC simulations.

\begin{figure*}[htb!]
 \plottwo{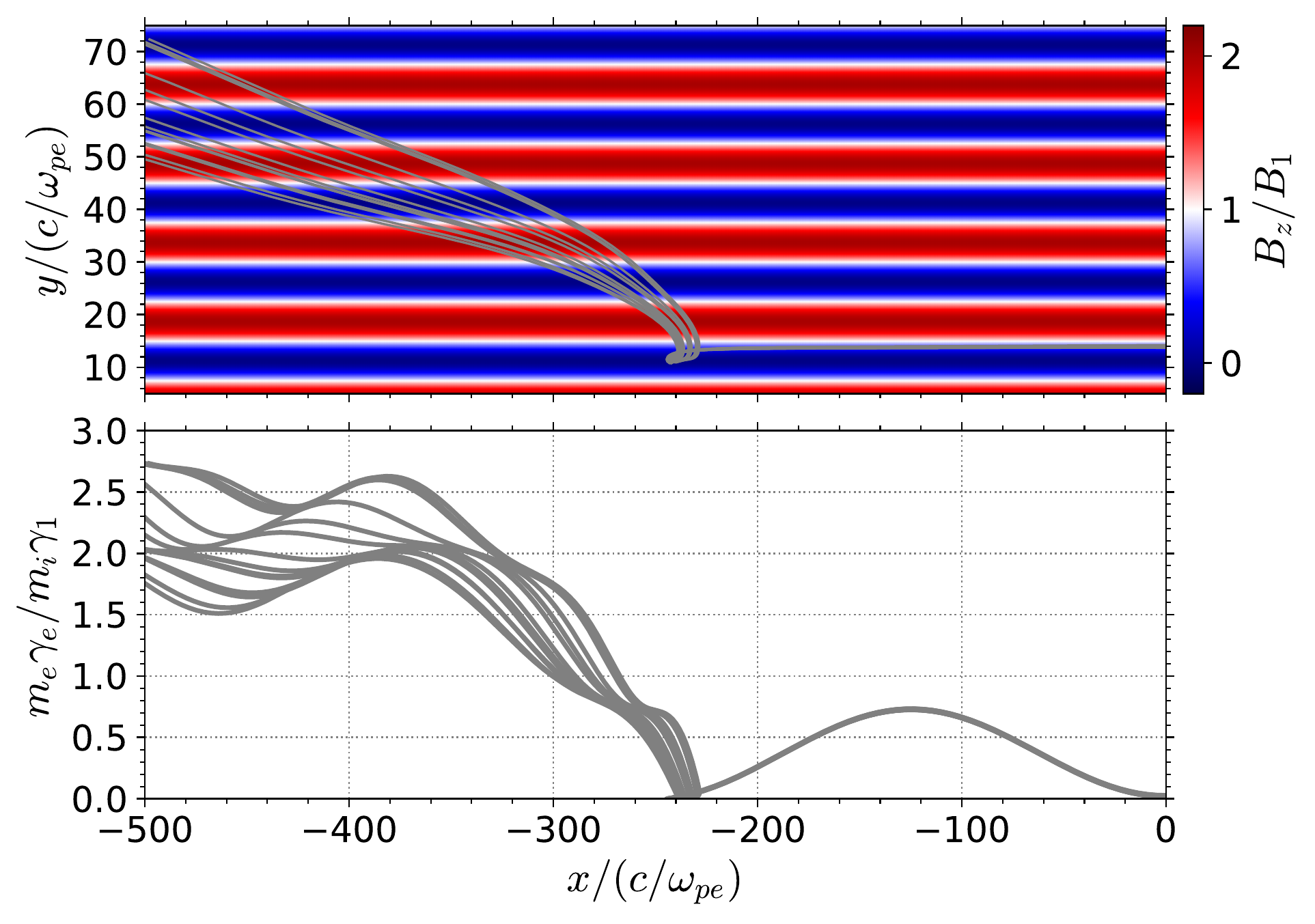}{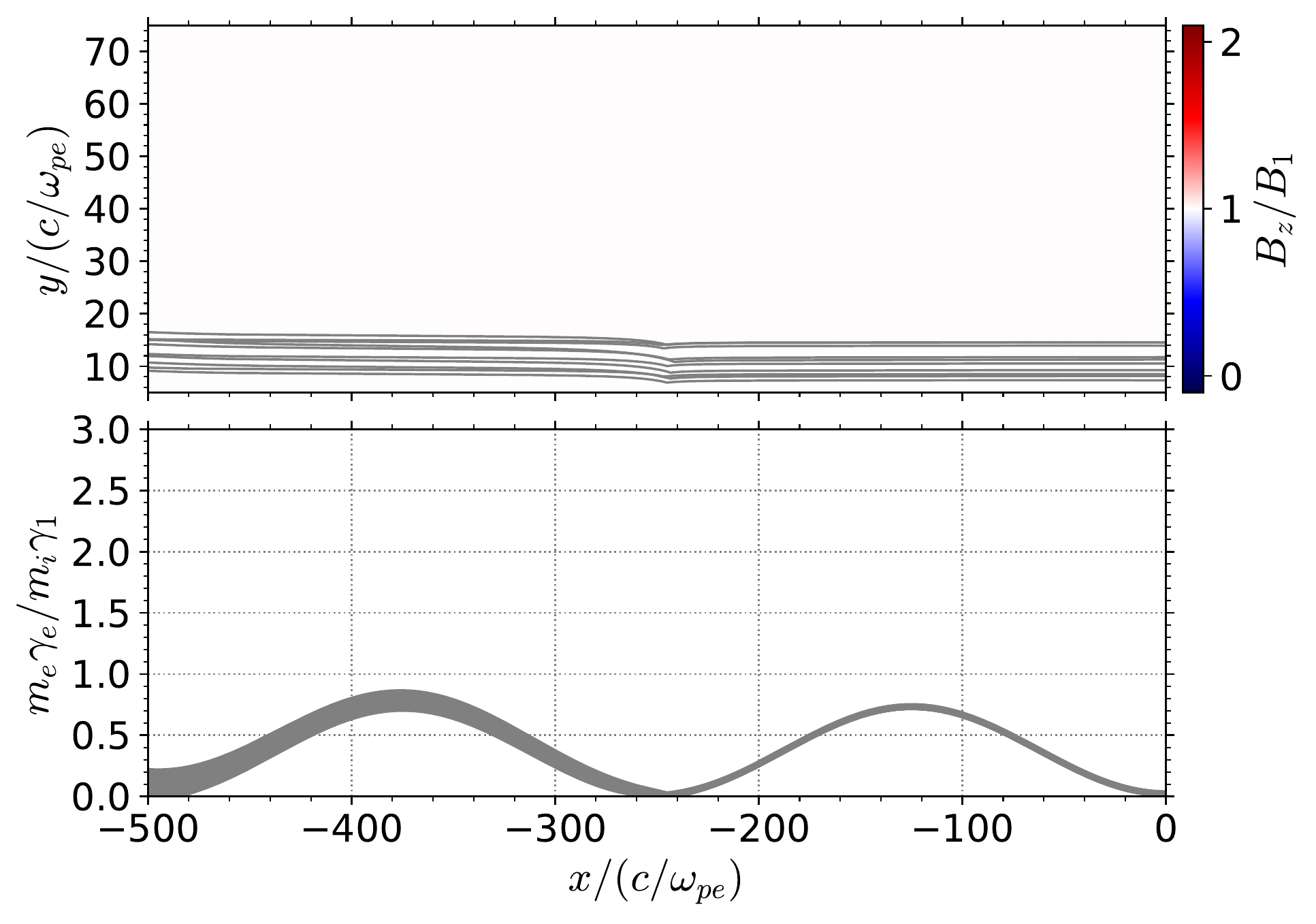}
 \caption{Electron trajectories with (left) and without (right) the filaments.}
 \label{fig:filament}
\end{figure*}

Figure \ref{fig:filament} show the trajectories of the energetic electrons in
$x$--$y$ space (top) and $x$--$\gamma$ space (bottom). The color maps represent
$B_z$ in the case of $B_f = 1$ (left) and $B_f = 0$ (right).
The incoming electrons are picked up at $x/(c/\omega_{pe}) \sim -220$ for
$B_f = 1$, whereas they are merely oscillating inside the wakefield for
$B_f = 0$. The trajectories give a clear proof
that the filaments are essential for entering the pickup process.

\begin{figure}[htb!]
 \plotone{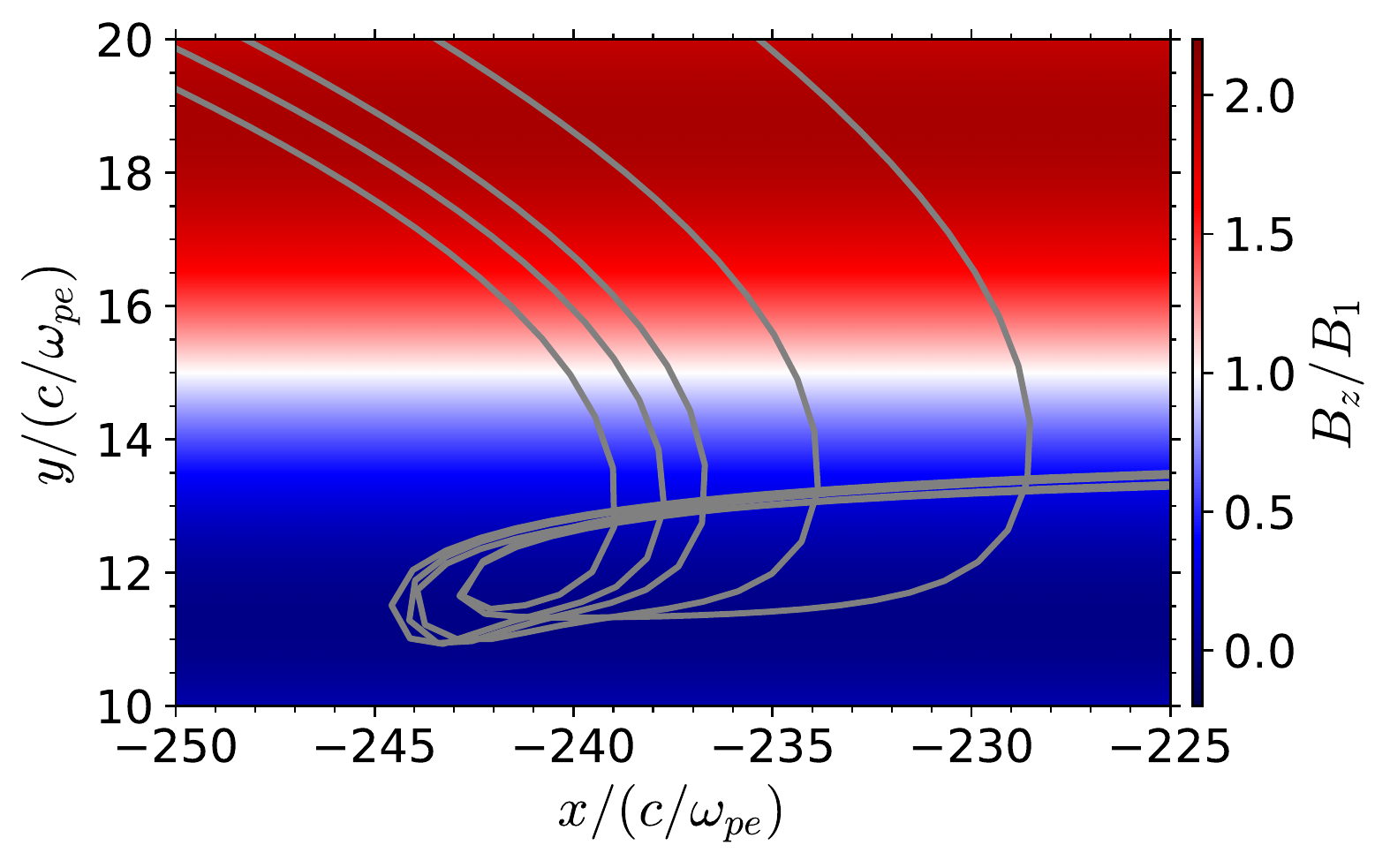}
 \caption{Trajectories of the energetic electrons with
 the magnetic field $B_z$.}
 \label{fig:bz_test}
\end{figure}

Figure \ref{fig:bz_test} displays the enlarged view of the top-left panel
of Figure \ref{fig:filament}.
The incoming electrons enter the unmagnetized region arising from the filaments
and then they are picked up by the bulk flow.
The filaments obviously triggers the pickup process as in the case with our
PIC simulations.

\begin{figure}[htb!]
 \plotone{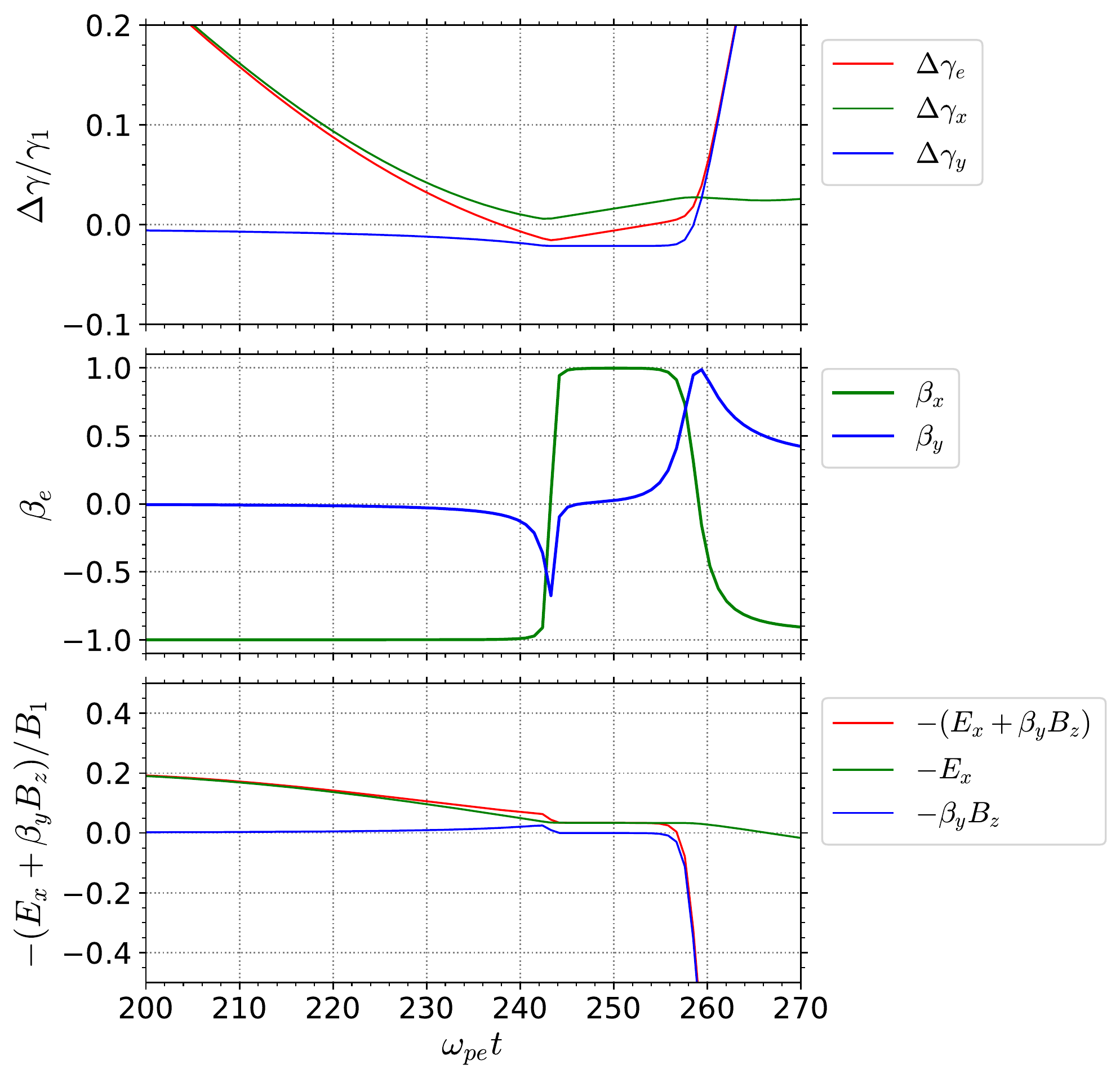}
 \caption{Time evolution of the typical energetic
 electron in the same format as Figure \ref{fig:evo_e}}
 \label{fig:evo_test}
\end{figure}

Figure \ref{fig:evo_test} shows the time evolution of the
typical energetic electron
in the same format as Figure \ref{fig:evo_e}.
The time evolution in our test particle simulation
exhibit the qualitatively same behavior as that in our PIC simulation.
The incoming electron is first decelerated by the wakefield and
begins to gyrate. The motional electric field
as well as the wakefield then decelerates it
for $210 \lesssim \omega_{pe}t \lesssim 244$.
The increase of $\Delta \gamma_x$ and the positive velocity
$\beta_x \sim 1$ for $244 \lesssim \omega_{pe}t \lesssim 256$
shows that the electrons are accelerated by the wakefield via the Landau resonance.
The electrostatic force dominates over the Lorentz force for
$244 \lesssim \omega_{pe}t \lesssim 256$ and thus the Landau resonance
can work.
After $\omega_{pe}t \simeq 256$,
$\Delta \gamma_y$ becomes dominant and the electron enters the
pickup process. The Lorentz force indeed exceeds
the electrostatic force after $\omega_{pe}t \simeq 256$.
The electron is detrapped from the wakefield and
then picked up by the bulk flow.

The injection into the pickup process
is well-described by this toy model.
We thus think that the filaments trigger the pickup process.

\begin{figure}[htb!]
 \plotone{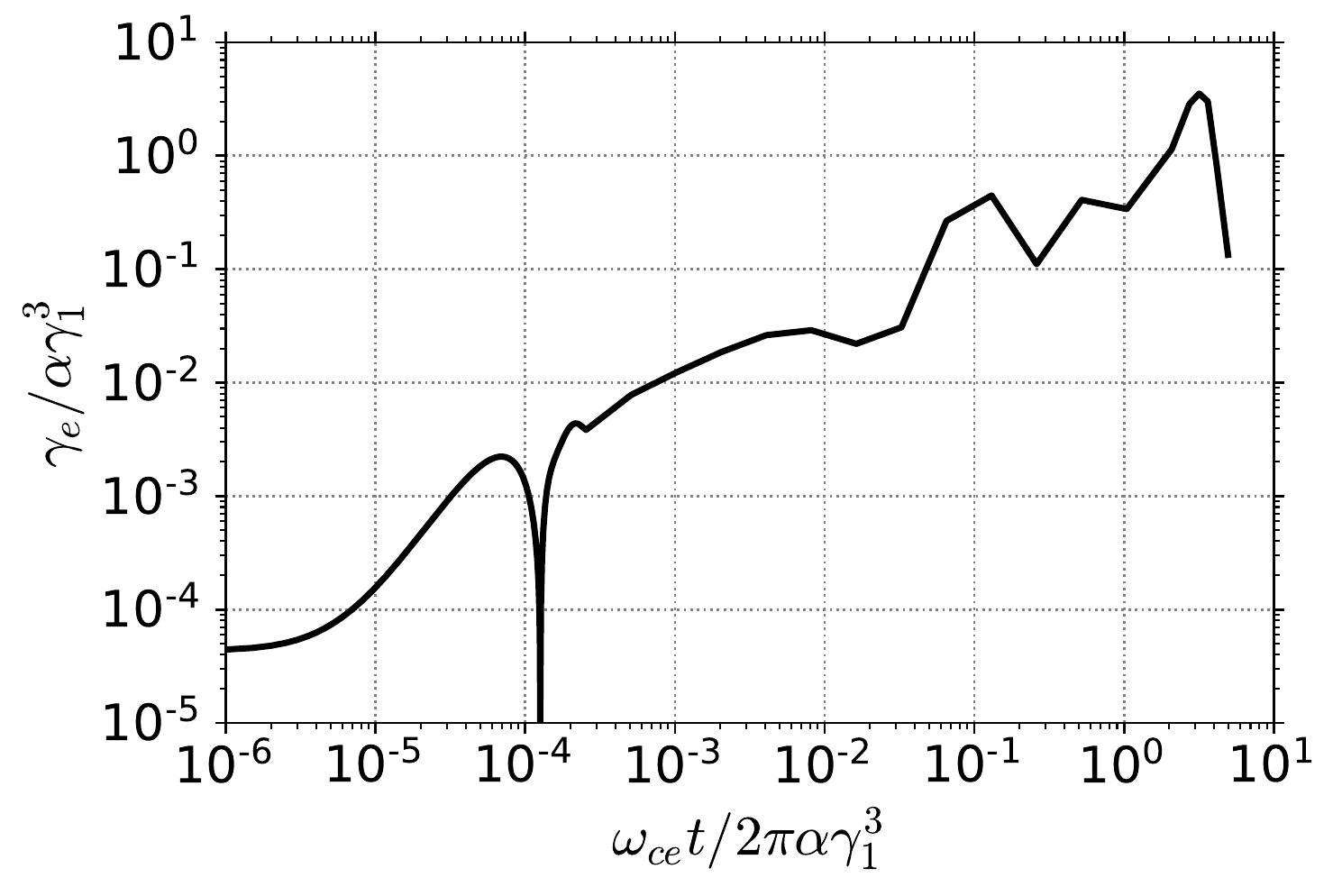}
 \caption{Time evolution of the Lorentz factor.}
 \label{fig:t-gam}
\end{figure}

Figure \ref{fig:t-gam} shows the time evolution of the ion Lorentz factor.
The maximum Lorentz factor is
$\gamma_{max,e} \sim \alpha \gamma_1^3$
and the acceleration timescale is $\omega_{ce}t \sim 2\pi\alpha\gamma_1^3$.
This result is consistent with the theoretical estimate discussed in the
main text, indicating that the pickup particles can be accelerated up to the
theoretical maximum Lorentz factor in the shock system if the size of the
precursor wave region is sufficiently large.

\bibliographystyle{aasjournal}
\bibliography{ref}

\begin{thebibliography}{}
\expandafter\ifx\csname natexlab\endcsname\relax\def\natexlab#1{#1}\fi
\providecommand{\url}[1]{\href{#1}{#1}}
\providecommand{\dodoi}[1]{doi:~\href{http://doi.org/#1}{\nolinkurl{#1}}}
\providecommand{\doeprint}[1]{\href{http://ascl.net/#1}{\nolinkurl{http://ascl.net/#1}}}
\providecommand{\doarXiv}[1]{\href{https://arxiv.org/abs/#1}{\nolinkurl{https://arxiv.org/abs/#1}}}

\bibitem[{Aab {et~al.}(2018)Aab, Abreu, Aglietta, Albuquerque, Allekotte,
  Almela, Castillo, Alvarez-Mu{\~{n}}iz, Anastasi, Anchordoqui,
  {et~al.}}]{Aab2018}
Aab, A., Abreu, P., Aglietta, M., {et~al.} 2018, \apjl, 853, L29,
  \dodoi{10.3847/2041-8213/aaa66d}

\bibitem[{Aartsen {et~al.}(2018)Aartsen, Ackermann, Adams, Aguilar, Ahlers,
  Ahrens, Samarai, Altmann, Andeen, Anderson, {et~al.}}]{Aartsen2018}
Aartsen, M.~G., Ackermann, M., Adams, J., {et~al.} 2018, Sci, 361, 147,
  \dodoi{10.1126/science.aat2890}

\bibitem[{Amato \& Arons(2006)}]{Amato2006}
Amato, E., \& Arons, J. 2006, \apj, 653, 325, \dodoi{10.1086/508050}

\bibitem[{Arons(2003)}]{Arons2003}
Arons, J. 2003, \apj, 589, 871, \dodoi{10.1086/374776}

\bibitem[{Babul \& Sironi(2020)}]{Babul2020}
Babul, A.-N., \& Sironi, L. 2020, \mnras, 499, 2884,
  \dodoi{10.1093/mnras/staa2612}

\bibitem[{Beloborodov(2017)}]{Beloborodov2017}
Beloborodov, A.~M. 2017, \apjl, 843, L26, \dodoi{10.3847/2041-8213/aa78f3}

\bibitem[{Beloborodov(2020)}]{Beloborodov2020}
---. 2020, \apj, 896, 142, \dodoi{10.3847/1538-4357/ab83eb}

\bibitem[{Blandford {et~al.}(2019)Blandford, Meier, \&
  Readhead}]{Blandford2019}
Blandford, R., Meier, D., \& Readhead, A. 2019, ARA\&A, 57, 467,
  \dodoi{10.1146/annurev-astro-081817-051948}

\bibitem[{Chen {et~al.}(2002)Chen, Tajima, \& Takahashi}]{Chen2002}
Chen, P., Tajima, T., \& Takahashi, Y. 2002, \prl, 89, 161101,
  \dodoi{10.1103/PhysRevLett.89.161101}

\bibitem[{Drake {et~al.}(1974)Drake, Kaw, Lee, Schmid, Liu, \&
  Rosenbluth}]{Drake1974}
Drake, J.~F., Kaw, P.~K., Lee, Y.~C., {et~al.} 1974, PhFl, 14, 778,
  \dodoi{10.1063/1.1694789}

\bibitem[{Ebisuzaki \& Tajima(2014)}]{Ebisuzaki2014}
Ebisuzaki, T., \& Tajima, T. 2014, APh, 56, 9,
  \dodoi{10.1016/j.astropartphys.2014.02.004}

\bibitem[{Ebisuzaki \& Tajima(2021)}]{Ebisuzaki2021}
---. 2021, APh, 128, 102567, \dodoi{10.1016/j.astropartphys.2021.102567}

\bibitem[{Fried(1959)}]{Fried1959}
Fried, B.~D. 1959, PhFl, 2, 337, \dodoi{10.1063/1.1705933}

\bibitem[{Gallant {et~al.}(1992)Gallant, Hoshino, Langdon, Arons, \&
  Max}]{Gallant1992}
Gallant, Y.~A., Hoshino, M., Langdon, A.~B., Arons, J., \& Max, C.~E. 1992,
  \apj, 391, 73, \dodoi{10.1086/171326}

\bibitem[{Hillas(1984)}]{Hillas1984}
Hillas, A.~M. 1984, ARA\&A, 22, 425,
  \dodoi{10.1146/annurev.aa.22.090184.002233}

\bibitem[{Hoshino(2008)}]{Hoshino2008}
Hoshino, M. 2008, \apj, 672, 940, \dodoi{10.1086/523665}

\bibitem[{Hoshino \& Arons(1991)}]{Hoshino1991}
Hoshino, M., \& Arons, J. 1991, PhFlB, 3, 818, \dodoi{10.1063/1.859877}

\bibitem[{Hoshino {et~al.}(1992)Hoshino, Arons, Gallant, \&
  Langdon}]{Hoshino1992}
Hoshino, M., Arons, J., Gallant, Y.~A., \& Langdon, A.~B. 1992, \apj, 390, 454,
  \dodoi{10.1086/171296}

\bibitem[{Hoshino \& Shimada(2002)}]{Hoshino2002}
Hoshino, M., \& Shimada, N. 2002, \apj, 572, 880, \dodoi{10.1086/340454}

\bibitem[{Ikeya \& Matsumoto(2015)}]{Ikeya2015}
Ikeya, N., \& Matsumoto, Y. 2015, PASJ, 67, 64, \dodoi{10.1093/pasj/psv052}

\bibitem[{Iwamoto {et~al.}(2017)Iwamoto, Amano, Hoshino, \&
  Matsumoto}]{Iwamoto2017}
Iwamoto, M., Amano, T., Hoshino, M., \& Matsumoto, Y. 2017, \apj, 840, 52,
  \dodoi{10.3847/1538-4357/aa6d6f}

\bibitem[{Iwamoto {et~al.}(2018)Iwamoto, Amano, Hoshino, \&
  Matsumoto}]{Iwamoto2018}
---. 2018, \apj, 858, 93, \dodoi{10.3847/1538-4357/aaba7a}

\bibitem[{Iwamoto {et~al.}(2019)Iwamoto, Amano, Hoshino, Matsumoto, Niemiec,
  Ligorini, Kobzar, \& Pohl}]{Iwamoto2019}
Iwamoto, M., Amano, T., Hoshino, M., {et~al.} 2019, \apjl, 883, L35,
  \dodoi{10.3847/2041-8213/ab4265}

\bibitem[{Kaw {et~al.}(1973)Kaw, Schmid, \& Wilcox}]{Kaw1973}
Kaw, P.~K., Schmid, G., \& Wilcox, T. 1973, PhFl, 16, 1522,
  \dodoi{10.1063/1.1694552}

\bibitem[{Kruer(1988)}]{Kruer1988}
Kruer, W.~L. 1988, The Physics of Laser Plasma Interactions, ed. D.~Pines
  (Boston: Addison-Wesley)

\bibitem[{Kuramitsu {et~al.}(2012)Kuramitsu, Sakawa, Hoshino, Chen, \&
  Takabe}]{Kuramitsu2012}
Kuramitsu, Y., Sakawa, Y., Hoshino, M., Chen, S.~H., \& Takabe, H. 2012, HEDP,
  8, 266, \dodoi{10.1016/j.hedp.2012.03.016}

\bibitem[{Kuramitsu {et~al.}(2008)Kuramitsu, Sakawa, Kato, Takabe, \&
  Hoshino}]{Kuramitsu2008}
Kuramitsu, Y., Sakawa, Y., Kato, T., Takabe, H., \& Hoshino, M. 2008, \apj,
  682, 113, \dodoi{10.1086/591247}

\bibitem[{Kuramitsu {et~al.}(2011{\natexlab{a}})Kuramitsu, Nakanii, Kondo,
  Sakawa, Mori, Miura, Tsuji, Kimura, Fukumochi, Kashihara,
  {et~al.}}]{Kuramitsu2011a}
Kuramitsu, Y., Nakanii, N., Kondo, K., {et~al.} 2011{\natexlab{a}}, PhPl, 18,
  010701, \dodoi{10.1063/1.3528434}

\bibitem[{Kuramitsu {et~al.}(2011{\natexlab{b}})Kuramitsu, Nakanii, Kondo,
  Sakawa, Mori, Miura, Tsuji, Kimura, Fukumochi, Kashihara,
  {et~al.}}]{Kuramitsu2011b}
---. 2011{\natexlab{b}}, PhRvE, 83, 026401, \dodoi{10.1103/PhysRevE.83.026401}

\bibitem[{Langdon {et~al.}(1988)Langdon, Arons, \& Max}]{Langdon1988}
Langdon, A.~B., Arons, J., \& Max, C.~E. 1988, \prl, 61, 779,
  \dodoi{10.1103/PhysRevLett.61.779}

\bibitem[{Ligorini {et~al.}(2021{\natexlab{a}})Ligorini, Niemiec, Kobzar,
  Iwamoto, Bohdan, Pohl, Matsumoto, Amano, Matsukiyo, Esaki, \&
  Hoshino}]{Ligorini2021a}
Ligorini, A., Niemiec, J., Kobzar, O., {et~al.} 2021{\natexlab{a}}, \mnras,
  501, 4837, \dodoi{10.1093/mnras/staa3901}

\bibitem[{Ligorini {et~al.}(2021{\natexlab{b}})Ligorini, Niemiec, Kobzar,
  Iwamoto, Bohdan, Pohl, Matsumoto, Amano, Matsukiyo, \&
  Hoshino}]{Ligorini2021b}
---. 2021{\natexlab{b}}, \mnras, 502, 5065, \dodoi{10.1093/mnras/stab220}

\bibitem[{Liu {et~al.}(2019)Liu, Isayama, Chen, \& Kuramitsu}]{Liu2019}
Liu, Y.~L., Isayama, S., Chen, S.~H., \& Kuramitsu, Y. 2019, HEDP, 31, 64,
  \dodoi{10.1016/j.hedp.2019.03.004}

\bibitem[{Liu {et~al.}(2018)Liu, Kuramitsu, Isayama, \& Chen}]{Liu2018}
Liu, Y.~L., Kuramitsu, Y., Isayama, S., \& Chen, S.~H. 2018, PhPl, 25, 013110,
  \dodoi{10.1063/1.5006325}

\bibitem[{Liu {et~al.}(2017)Liu, Kuramitsu, Moritaka, \& Chen}]{Liu2017}
Liu, Y.~L., Kuramitsu, Y., Moritaka, T., \& Chen, S.~H. 2017, HEDP, 22, 46,
  \dodoi{10.1016/j.hedp.2017.02.006}

\bibitem[{Lyubarsky(2006)}]{Lyubarsky2006}
Lyubarsky, Y. 2006, \apj, 652, 1297, \dodoi{10.1086/508606}

\bibitem[{Lyubarsky(2008)}]{Lyubarsky2008}
---. 2008, \apj, 682, 1443, \dodoi{10.1086/589435}

\bibitem[{Lyubarsky(2014)}]{Lyubarsky2014}
---. 2014, \mnras, 442, L9, \dodoi{10.1093/mnrasl/slu046}

\bibitem[{Lyubarsky(2018)}]{Lyubarsky2018}
---. 2018, \mnras, 474, 1135, \dodoi{10.1093/mnras/stx2832}

\bibitem[{Lyubarsky(2019)}]{Lyubarsky2019}
---. 2019, \mnras, 490, 1474, \dodoi{10.1093/mnras/stz2712}

\bibitem[{Margalit {et~al.}(2020)Margalit, Beniamini, Sridhar, \&
  Metzger}]{Margalit2020}
Margalit, B., Beniamini, P., Sridhar, N., \& Metzger, B.~D. 2020, \apjl, 899,
  L27, \dodoi{10.3847/2041-8213/abac57}

\bibitem[{Matsumoto {et~al.}(2013)Matsumoto, Amano, \& Hoshino}]{Matsumoto2013}
Matsumoto, Y., Amano, T., \& Hoshino, M. 2013, \prl, 111, 215003,
  \dodoi{10.1103/PhysRevLett.111.215003}

\bibitem[{Matsumoto {et~al.}(2015)Matsumoto, Amano, Kato, \&
  Hoshino}]{Matsumoto2015}
Matsumoto, Y., Amano, T., Kato, T.~N., \& Hoshino, M. 2015, Sci, 347, 974,
  \dodoi{10.1126/science.1260168}

\bibitem[{Max {et~al.}(1974)Max, Arons, \& Langdon}]{Max1974}
Max, C.~E., Arons, J., \& Langdon, A.~B. 1974, \prl, 33, 209,
  \dodoi{10.1103/PhysRevLett.33.209}

\bibitem[{Melrose(2017)}]{Melrose2017}
Melrose, D.~B. 2017, RvMPP, 1, 5, \dodoi{10.1007/s41614-017-0007-0}

\bibitem[{Metzger {et~al.}(2019)Metzger, Margalit, \& Sironi}]{Metzger2019}
Metzger, B.~D., Margalit, B., \& Sironi, L. 2019, \mnras, 485, 4091,
  \dodoi{10.1093/mnras/stz700}

\bibitem[{M{\"{o}}bius {et~al.}(1985)M{\"{o}}bius, Hovestadt, Klecker, Scholer,
  Gloeckler, \& Ipavich}]{Mobius1985}
M{\"{o}}bius, E., Hovestadt, D., Klecker, B., {et~al.} 1985, Nature, 318, 426,
  \dodoi{10.1038/318426a0}

\bibitem[{Murase {et~al.}(2009)Murase, M{\'{e}}sz{\'{a}}ros, \&
  Zhang}]{Murase2009}
Murase, K., M{\'{e}}sz{\'{a}}ros, P., \& Zhang, B. 2009, PhRvD, 79, 103001,
  \dodoi{10.1103/PhysRevD.79.103001}

\bibitem[{Oka {et~al.}(2002)Oka, Terasawa, Noda, Saito, \& Mukai}]{Oka2002}
Oka, M., Terasawa, T., Noda, H., Saito, Y., \& Mukai, T. 2002, GeoRL, 29, 1612,
  \dodoi{10.1029/2002GL015111}

\bibitem[{Piran(2005)}]{Piran2005}
Piran, T. 2005, RvMP, 76, 1143, \dodoi{10.1103/RevModPhys.76.1143}

\bibitem[{Plotnikov {et~al.}(2018)Plotnikov, Grassi, \& Grech}]{Plotnikov2018}
Plotnikov, I., Grassi, A., \& Grech, M. 2018, \mnras, 477, 5238,
  \dodoi{10.1093/mnras/sty979}

\bibitem[{Plotnikov \& Sironi(2019)}]{Plotnikov2019}
Plotnikov, I., \& Sironi, L. 2019, \mnras, 485, 3816,
  \dodoi{10.1093/mnras/stz640}

\bibitem[{Shimada \& Hoshino(2000)}]{Shimada2000}
Shimada, N., \& Hoshino, M. 2000, \apjl, 543, L67, \dodoi{10.1086/318161}

\bibitem[{Sironi {et~al.}(2021)Sironi, Plotnikov, N{\"{a}}ttil{\"{a}}, \&
  Beloborodov}]{Sironi2021}
Sironi, L., Plotnikov, I., N{\"{a}}ttil{\"{a}}, J., \& Beloborodov, A.~M. 2021,
  \prl, 127, 035101, \dodoi{10.1103/PhysRevLett.127.035101}

\bibitem[{Sironi \& Spitkovsky(2011)}]{Sironi2011}
Sironi, L., \& Spitkovsky, A. 2011, \apj, 726, 75,
  \dodoi{10.1088/0004-637X/726/2/75}

\bibitem[{Sironi {et~al.}(2013)Sironi, Spitkovsky, \& Arons}]{Sironi2013}
Sironi, L., Spitkovsky, A., \& Arons, J. 2013, \apj, 771, 54,
  \dodoi{10.1088/0004-637X/771/1/54}

\bibitem[{Sobacchi {et~al.}(2020)Sobacchi, Lyubarsky, Beloborodov, \&
  Sironi}]{Sobacchi2020}
Sobacchi, E., Lyubarsky, Y., Beloborodov, A.~M., \& Sironi, L. 2020, \mnras,
  500, 272, \dodoi{10.1093/mnras/staa3248}

\bibitem[{Spitkovsky(2005)}]{Spitkovsky2005}
Spitkovsky, A. 2005, in AIP Conf. Proc., Vol. 801, Astrophysical Source of High
  Energy Particles and Radiation, ed. T.~Bulik, B.~Rudak, \& G.~Madejski
  (Melville, NY: AIP), 345--350, \dodoi{10.1063/1.2141897}

\bibitem[{Tajima \& Dawson(1979)}]{Tajima1979}
Tajima, T., \& Dawson, J.~M. 1979, \prl, 43, 267,
  \dodoi{10.1103/PhysRevLett.43.267}

\bibitem[{Vay(2008)}]{Vay2008}
Vay, J.-L. 2008, PhPl, 15, 056701, \dodoi{10.1063/1.2837054}

\bibitem[{Weibel(1959)}]{Weibel1959}
Weibel, Erich, S. 1959, \prl, 2, 83, \dodoi{10.1103/PhysRevLett.2.83}

\end{thebibliography}

\listofchanges
\end{document}